\documentclass[11pt,a4paper]{article}
\usepackage{times,amssymb,cite,bm}
\newcommand{\be}{\begin{eqnarray}}
\newcommand{\ee}{\end{eqnarray}}
\newcommand{\bez}{\begin{eqnarray*}}
\newcommand{\eez}{\end{eqnarray*}}
\newcommand{\pa}{\partial}
\newcommand{\la}{\lambda}
\newcommand{\A}{\mathbb{A}}
\newcommand{\C}{\mathbb{C}}
\newcommand{\K}{\mathbb{K}}
\newcommand{\R}{\mathbb{R}}
\newcommand{\Cdot}{\boldsymbol{\cdot}}
\newcommand{\hH}{\hat{H}}
\newcommand{\hE}{\hat{E}}

\newcommand{\ci}{\circ}
\newcommand{\tr}{\mathrm{tr}}
\newcommand{\bfalpha}{\mbox{\boldmath$\alpha$}}
\newcommand{\bfbeta}{\mbox{\boldmath$\beta$}}
\newcommand{\bfgamma}{\mbox{\boldmath$\gamma$}}
\newcommand{\sbfalpha}{\mbox{\boldmath$\scriptstyle \alpha$}}
\newcommand{\sbfbeta}{\mbox{\boldmath$\scriptstyle \beta$}}
\newcommand{\sbfgamma}{\mbox{\boldmath$\scriptstyle \gamma$}}
\newcommand{\bq}{\bar{q}}
\newcommand{\bpsi}{\bar{\psi}}
\newcommand{\bmu}{\bar{\mu}}

\title{\bf Nonassociativity and Integrable Hierarchies
\thanks{\copyright 2006 by A. Dimakis and F. M\"uller-Hoissen} }

\author{Aristophanes Dimakis \\
 Department of Financial and Management Engineering, \\
 University of the Aegean, 31 Fostini Str., GR-82100 Chios, Greece \\
 dimakis@aegean.gr
          \and
 Folkert M\"uller-Hoissen \\ Max-Planck-Institute for Dynamics and Self-Organization \\
 Bunsenstrasse 10, D-37073 G\"ottingen, Germany \\
 folkert.mueller-hoissen@ds.mpg.de }

\date{}

\begin{document}

\renewcommand{\theequation} {\arabic{section}.\arabic{equation}}

\newtheorem{theorem}{Theorem}[section]
\newtheorem{lemma}[theorem]{Lemma}
\newtheorem{proposition}[theorem]{Proposition}
\newtheorem{definition}{Definition}[section]
\newtheorem{corollary}[theorem]{Corollary}

\maketitle

\begin{abstract}
Let $\A$ be a nonassociative algebra such that the associator $(\A,\A^2,\A)$ vanishes. 
If $\A$ is freely generated by an element $f$, there are commuting derivations 
$\delta_n$, $n=1,2, \ldots$, such that 
$\delta_n(f)$ is a nonlinear homogeneous polynomial in $f$ of degree $n+1$. 
We prove that the expressions $\delta_{n_1} \cdots \delta_{n_k}(f)$ satisfy 
identities which are in correspondence with the equations of the 
Kadomtsev-Petviashvili (KP) hierarchy. As a consequence, solutions of the 
`nonassociative hierarchy' $\pa_{t_n}(f) = \delta_n(f)$, $n=1,2,\ldots$, of 
ordinary differential equations lead to solutions of the KP hierarchy. 
The framework is extended by introducing the notion of an $\A$-module and 
constructing, with the help of the derivations $\delta_n$, zero curvature 
connections and linear systems. 
\end{abstract}

\small
\tableofcontents
\normalsize

\section{Introduction}
Let $f$ generate freely a nonassociative algebra $\A$ over a commutative 
ring $\mathcal{R}$ with identity element (see \cite{Bour89,Kuz'+Shes95}, 
for example, for the algebraic structures used in this work). 
In order to construct commuting derivations, we choose their actions on the 
generator as nonlinear homogeneous expressions in the generator and extend 
them via the derivation rule. 
For the first equation (of lowest order in $f$), there is no other choice than
\be
    \delta_1(f) = f^2  \; .     \label{hier1}
\ee
The second equation should take the form
\be
   \delta_2(f) = \kappa_1 \, f f^2 - \kappa_2 \, f^2 f      \label{hier2}
\ee
with $\kappa_1, \kappa_2 \in \mathcal{R}$, since $f f^2$ and $f^2 f$ are the only 
independent monomials cubic in $f$. The requirement that $\delta_1$ and 
$\delta_2$ are commuting derivations then leads to the condition
\be
   (\kappa_1 + \kappa_2) \, (f,f^2,f)  = (\kappa_1 - \kappa_2) \, f^2 f^2 \, ,
\ee
with the associator
\be
    (a,b,c) := (a b) \, c - a \, (b c) \; .
\ee 
Choosing $\kappa_1 = \kappa_2 =1$, this means
\be
       (f,f^2,f) = 0 \, , 
\ee
which weakens the allowed nonassociativity. In fact, we will more generally 
assume that
\be
    (a,b \, c,d)
  \equiv (a \, (b \, c)) \, d - a \, ((b \, c) \, d) = 0
                             \label{weak_nonassociativity}
\ee
for all $a, b, c, d \in \A$. 
The next derivation can then be taken as
\be
    \delta_3(f) = f (f f^2) - f f^2 f - f^2 f^2 + (f^2 f) f \; .  \label{hier3}
\ee
This construction can consistently be continued ad infinitum (note that 
derivations preserve (\ref{weak_nonassociativity})), 
and the underlying general building law will be presented in this 
work. The condition (\ref{weak_nonassociativity}) is a rather 
strong restriction of the a priori possible nonassociativity. In this work 
we will not address the problem of finding the weakest restriction of 
nonassociativity and the corresponding hierarchy of derivations. 
\vskip.2cm

The derivations $\delta_n$ are subject to algebraic 
identities. For example, a direct calculation reveals that
\be
   \delta_1 \Big( 4 \, \delta_3(f) - \delta_1^3(f) + 6 \, (\delta_1(f))^2 \Big)
   - 3 \, \delta_2^2(f) + 6 \, [\delta_2(f) , \delta_1(f) ] \equiv 0
   \label{KP-id}
\ee
as a consequence of the definitions (\ref{hier1}), (\ref{hier2}), 
(\ref{hier3}), and the derivation rule. 
If we formally replace the $\delta_n$ by partial derivatives $\pa_{t_n}$ 
with respect to independent variables $t_1,t_2, \ldots$, 
we recover the potential KP equation. 
In fact, the elements $\delta_{n_1} \cdots \delta_{n_k}(f)$, 
where $n_1, \ldots, n_k = 1,2, \ldots$ and $k=1,2, \ldots$, satisfy more 
identities of this kind and, as we will prove in this work, the whole 
KP hierarchy emerges in this way. 
Since these identities are built solely from the `composite elements' 
$\delta_{n_1} \cdots \delta_{n_k}(f)$, 
as a consequence of (\ref{weak_nonassociativity}) they actually live in an 
\emph{associative} subalgebra. 
\vskip.2cm

If $\A$ is taken over a commutative ring of (smooth) functions of  
independent variables $t_1,t_2,\ldots$, we may consider the hierarchy
\be
   \pa_{t_n}(f) = \delta_n(f)  \qquad\quad  n = 1,2,\ldots
                              \label{naKP_hier}
\ee
of commuting flows in the nonassociative algebra $\A$.
This turns the identity (\ref{KP-id}) into the potential KP equation. 
According to the more general result mentioned above, the whole potential 
KP hierarchy is a consequence of (\ref{naKP_hier}). 
This implies that any solution of (\ref{naKP_hier}) leads to a solution 
of the potential KP hierarchy. Then $u := \pa_{t_1} f$, which as a 
consequence of (\ref{hier1}), (\ref{naKP_hier}) and (\ref{weak_nonassociativity}) 
lies in an \emph{associative} subalgebra, solves the KP hierarchy. 
\vskip.2cm

Since (\ref{naKP_hier}) is an autonomous system of first order differential 
equations with commuting flows, in certain cases it has a (formal) solution 
for given initial data. We will show that, with a suitable choice of $\A$, 
from this solution one obtains in particular solutions of the (potential) 
KP hierarchy with dependent variable in a matrix algebra, 
which in turn lead to familiar solutions of the scalar potential KP hierarchy, 
including the multi-solitons (see section~\ref{section:solutions}).
\vskip.2cm

The right hand sides of the equations (\ref{naKP_hier}) are purely 
algebraic expressions in $f$ and do not involve derivatives. They display 
the combinatorial structure underlying the KP (and also Burgers) hierarchy. 
The first equation $f_{t_1} = f^2$ of the hierarchy (\ref{naKP_hier}), 
which is the only one that would survive in case of associativity, 
is the equation of a `nonassociative top' \cite{GSS93}. It has the 
form of (nonassociative) `quadratic dynamical systems' as considered in 
\cite{Kiny+Sagl95,Kiny+Sagl02}, for example (see also the references therein). 
\vskip.2cm

In a different way, nonassociative algebras already made their appearance in 
the context of integrable systems 
\cite{Svin92,Svin94,Svin+Soko94,Soko+Svin95,Svin+Soko96,Golu+Soko00,Stra00,Adle00,GKT01,Kiny+Sagl02}. 
Svinolupov \cite{Svin92} determined the conditions under which a system of 
equations of a special form, involving `structure constants' of a nonassociative 
algebra, possesses symmetries, i.e., other systems such that the flows commute. 
This led to algebraic structures known as `left-symmetric algebras' 
(see also \cite{Bai04} and references therein) and `Jordan pairs'. 
\vskip.2cm

In section~\ref{section:na} we introduce the basic nonassociative 
algebraic structure on which this work is rooted. 
Section~\ref{section:products} introduces a sequence of 
derived products in any algebra which is `weakly nonassociative' in the sense 
of (\ref{weak_nonassociativity}). This mainly serves 
as a preparation for the construction in section~\ref{section:derivations} 
of a hierarchy of derivations, i.e. a sequence of commuting derivations 
$\delta_n$, $n=1,2, \ldots$, for a subclass of weakly nonassociative algebras. Section~\ref{section:KPids} then contains a major result of this work, 
namely the proof that the derivations $\delta_n$ satisfy a sequence of 
algebraic identities which are in correspondence with the equations of 
the KP hierarchy (as outlined above). 
An application of this result to quasi-symmetric functions is given 
in section~\ref{section:qs}. 
In section~\ref{section:nahier} we derive some properties and consequences of 
the `nonassociative hierarchy' (\ref{naKP_hier}). 
Section~\ref{section:solutions} demonstrates in particular how multi-soliton 
solutions of the (potential) KP hierarchy are recovered in this framework. 
In section~\ref{section:A-modules} we extend the algebraic 
framework by introducing (left- and right-) $\A$-modules and connections on
them. Section~\ref{section:concl} contains some concluding remarks.

\section{Weakly nonassociative algebras}
\label{section:na}
\setcounter{equation}{0}
Let $\A$ be any (nonassociative) algebra over a unital commutative ring $\mathcal{R}$. 
Then 
\be
    \A' := \{ b \in \A \, | \, (a,b,c) = 0  \quad \forall a,c \in \A \}
\ee
is an \emph{associative} subalgebra. If $\A$ has an identity element, 
then it belongs to $\A'$. 
We call $\A$ \emph{weakly nonassociative (WNA)} if $\A^2 \subset \A'$, which 
is equivalent to the condition
\be
    (a, b c, d) = 0 \qquad \quad \forall a,b,c,d \in \A \; .  \label{weak_nonassoc}
\ee
In this case $\A'$ is also a two-sided ideal in $\A$. 
Obviously, for a WNA algebra $\A$ the quotient algebra $\A/\A'$ is nilpotent of index 2. 
The condition (\ref{weak_nonassoc}) can also be expressed in the following two ways,
\be
        \begin{array}{l@{\, = \,}l}
        L_a \, L_b  & L_{a b} \\
        R_a \, R_b & R_{b a}
	\end{array}  \qquad \mbox{if} \quad b \in \A'  \, , 
	\label{LR(prod)}
\ee
where $L_a, R_a$ are left and right multiplication by $a \in \A$. 
\vskip.2cm

If $\A$ is any algebra and $\mathcal{I}$ the two-sided ideal in $\A$ 
generated by $(a, b c, d)$ for all $a,b,c,d \in \A$, then $\A/\mathcal{I}$ 
is a WNA algebra. 
\vskip.2cm

For a WNA algebra $\A$, let $\A(f)$ denote the subalgebra generated by a 
single element $f \in \A$. We have $L_f R_f(f) = L_f(f^2)$, $R_f L_f(f) = R_f(f^2)$, 
and $L_f R_f(b) = R_f L_f(b)$ if $b \in \A'$. 
Since $R_f^m(a) \in \A'$ and $L_f^m(a) \in \A'$ for $m \geq 1$ and 
for all $a \in \A$, the following relations hold, 
\be
    L_f^n R_f^{m+1} = R_f^m L_f^n R_f \, ,  \qquad
    R_f^n L_f^{m+1} = L_f^m R_f^n L_f \qquad \quad  m,n = 0,1,2, \ldots \; .
    \label{LR-rels}
\ee
In particular, this implies
\be
    L_f^n R_f^{m+1}(f) = L_f^n R_f^m(f^2) = R_f^m L_f^{n+1}(f)
    \qquad \quad      m,n = 0,1,2, \ldots \; .  \label{LR-RL}
\ee
It is convenient to introduce the abbreviations
\be
   h_n := L_f^n(f)     \, , \qquad
   e_n := R_f^n(f)     \qquad \quad
   n=0,1,2, \ldots    \; .    \label{h_n&e_n}
\ee
The reader should keep in mind that $h_n$, $e_n$, and objects expressed 
in terms of them, depend on the choice of $f$. 
For the sake of a simpler notation, we do not write this dependence explicitly.

\begin{proposition}
\label{prop:span}
The algebra $\A(f)$ is spanned by $f$ and products of the elements 
$L_f^n R_f^m(f^2)$, $n,m=0,1,2,\ldots$.
\end{proposition}
{\it Proof:}
Obviously, $\A(f)$ is spanned by products of monomials of the form 
$L_f^{i_m} R_f^{j_n} \cdots L_f^{i_1} R_f^{j_1}(f)$ where $i_1,\ldots,i_m, j_1,
\ldots, j_n \in \mathbb{N} \cup \{0\}$ and $m,n \in \mathbb{N}$. 
Using (\ref{LR-rels}) and (\ref{LR-RL}), we see that such monomials are 
given by $f$ or $L_f^n R_f^m(f^2)$, where $m,n \geq 0$. 
As a consequence of the WNA property, 
any monomial built from elements of the form $L_f^r R_f^s(f^2)$ and also 
with $f$ can be reduced to a monomial which consists of products of elements 
$L_f^n R_f^m(f^2)$ only.
\hfill $\square$
\vskip.2cm

\noindent
{\bf Example.} Let $\mathcal{A}$ be an \emph{associative} algebra and 
$L_i, R_i \, : \, \mathcal{A} \to \mathcal{A}$, $i=1,2,\ldots,N$, 
linear maps such that $[L_i,R_j]=0$ and 
$L_i(a b) = L_i(a) \, b, R_i(a b)= a \, R_i(b)$ for all $a,b \in \mathcal{A}$. 
Then we can construct a WNA algebra $\A$ by augmenting $\mathcal{A}$ 
with elements $f_i$, $i=1,\ldots,N$, such that 
\be
       f_i f_j = g_{ij} \, , 
\ee
with fixed $g_{ij} \in \mathcal{A}$, and setting
\be
    f_i \, a := L_i(a) \, , \qquad 
    a \, f_i := R_i(a)  \qquad \quad 
           \forall a \in \mathcal{A} \; .
\ee
Obviously, $\mathcal{A} \subset \A'$. In fact, $\A'$ will be larger than $\mathcal{A}$ 
if the extension by some $f_k$ is associative, a case which is of less interest. 
For fixed $k$, $f_k$ lies in $\A'$ if and only if
\be
   L_i L_k(a) = g_{ik} \, a \, , \quad
   R_i R_k(a) = a \, g_{ki} \, , \quad
   R_j(g_{ik}) = L_i(g_{kj}) \, , \quad
   R_k(a) \, b = a \, L_k(b) \, , 
\ee
for $i,j=1,\ldots,N$ and for all $a,b \in \mathcal{A}$. 
A precise formulation of the above augmentation is given in the next proposition. 
Particular examples are obtained by setting 
$L_i(a) = q_i \, a$, $R_i(a) = a \, r_i$, with fixed elements $q_i, r_i \in \mathcal{A}$. 
If $\mathcal{A}$ has an identity element (unity) $u$, then $q_i = L_i(u)$ and 
$r_i = R_i(u)$.
\hfill $\square$

\begin{proposition}
\label{prop:WNA}
Let $\mathcal{A}$ be an associative algebra (over $\mathcal{R}$), 
$L_i, R_i \, : \, \mathcal{A} \rightarrow \mathcal{A}$, $i=1,\ldots,N$, 
linear maps such that $[L_i,R_j]=0$ and 
\be
    L_i(a b) = L_i(a) \, b \, , \qquad R_i(a b) = a \, R_i(b) 
      \qquad \quad \forall a,b \in \mathcal{A}    \, ,    \label{L_i,R_i-actions}
\ee
and $g_{ij} \in \A$, $i,j = 1,\ldots,N$. For 
$\bfalpha = (\alpha_1,\ldots,\alpha_N), \bfbeta = (\beta_1,\ldots,\beta_N) \in \mathcal{R}^N$, 
let us define 
\be
    g(\bfalpha,\bfbeta) := \sum_{i,j=1}^N \alpha_i \beta_j \, g_{ij} \, , \qquad
    L_{\sbfalpha} := \sum_{i=1}^N \alpha_i \, L_i \, , \qquad
    R_{\sbfbeta} := \sum_{i=1}^N \beta_i \, R_i \; .
\ee
Then $\A := (\bigoplus_{i=1}^N \mathcal{R}) \oplus \mathcal{A}$, 
supplied with the product
\be
    (\bfalpha,a)(\bfbeta,b) 
 := (\mathbf{0},g(\bfalpha,\bfbeta) + L_{\sbfalpha}(b) + R_{\sbfbeta}(a) + a b) \, ,
    \label{canonical-WNA-product}
\ee
is a WNA algebra. 
If, in addition, the equations 
\be
    L_{\sbfbeta} L_{\sbfalpha}(a) = g(\bfbeta,\bfalpha) \, a  , \; 
    R_{\sbfbeta} R_{\sbfalpha}(a) = a \, g(\bfalpha,\bfbeta)  , \; 
    R_{\sbfgamma}(g(\bfbeta,\bfalpha)) = L_{\sbfbeta}(g(\bfalpha,\bfgamma)) , \; 
    R_{\sbfalpha}(a) \, b = a \, L_{\sbfalpha}(b) , \;\;   \label{na-gen-cond}
\ee
for all $a,b \in \mathcal{A}$ and all $\bfbeta,\bfgamma \in \mathcal{R}^N$,
imply $\bfalpha = \mathbf{0}$, then $\A' = \mathcal{A}$, and $\A/\A'$ is a free 
module with $N$ generators. 
\\
Furthermore, any WNA algebra $\A$, for which $\A/\A'$ is finitely generated, 
is isomorphic to a WNA algebra of this type. 
\end{proposition}
{\it Proof:} It is easily verified that the algebra obtained by the above construction 
indeed satisfies (\ref{weak_nonassoc}) (see also the last example). 
The additional condition then guarantees that 
$f_i := (0,\ldots,1,0,\ldots,0,0) \not\in \A'$ (with the identity element of 
$\mathcal{R}$ at the $i$th position) for $i=1,\ldots,N$, 
and that the equivalence classes $[f_i] \in \A/\A'$, $i=1,\ldots,N$, are independent. 

Conversely, let $\A$ be a WNA algebra and $f_i$, $i=1,\ldots,N$, elements such that 
$[f_i]$, $i=1,\ldots,N$, freely generate $\A/\A'$. 
Then $g_{ij} := f_i f_j \in \A'$, and $L_i(a) := f_i \, a$, 
$R_i(a) := a \, f_i$ define linear maps $\A' \rightarrow \A'$. 
As a consequence of the WNA property, we have $[L_i,R_j]=0$ and (\ref{L_i,R_i-actions}) 
with $\mathcal{A} := \A'$. 
The set of equations (\ref{na-gen-cond}) is then equivalent to  
\bez
 && \sum_{i,j=1}^N \beta_i \alpha_j \, (f_i,f_j,a) = 0 \, , \quad
    \sum_{i,j=1}^N \alpha_i \beta_j \, (a,f_i,f_j) = 0 \, , \quad  \\
 && \sum_{i,j,k=1}^N \beta_i \alpha_j \gamma_k \, (f_i,f_j,f_k) = 0  \, , \quad
    \sum_{i=1}^N \alpha_i \, (a,f_i,b) = 0  \, , 
\eez
for all $a,b \in \A'$, and for all $\bfbeta,\bfgamma \in \mathcal{R}^N$. 
Since $[f_i]$, $i=1,\ldots,N$, are independent, this implies $\bfalpha =0$. 
It is easily verified that $\iota(a) := (\mathbf{0},a)$ for all  $a \in \A'$, 
and $\iota(f_i) := (0,\ldots,1,0,\ldots,0,0)$ (with the identity element of 
$\mathcal{R}$ at the $i$th position), $i=1,\ldots, N$, determines an isomorphism 
$\iota : \A \rightarrow (\bigoplus_{i=1}^N \mathcal{R}) \oplus A'$, where the target 
is supplied with the product (\ref{canonical-WNA-product}). 
\hfill $\square$
\vskip.2cm

If not stated otherwise, in the following $\A$ will always refer to a 
nonassociative (and typically noncommutative) algebra which is WNA. 
Occasionally we need to extend the ring $\mathcal{R}$, over which the algebra 
$\A$ is taken, to a ring of formal power series in certain parameters. 
This will not always be mentioned explicitly.

\subsection{The free WNA algebra with a single generator}
\label{subsection:freeWNA}
Let $\mathcal{A}_{\mathrm{free}}$ be the \emph{free associative} 
algebra over $\mathcal{R}$, generated by elements $c_{m,n}$, $m,n=0,1,\ldots$. 
We define linear maps 
$L,R : \mathcal{A}_{\mathrm{free}} \to \mathcal{A}_{\mathrm{free}}$ by
\be
    L(c_{m,n}) := c_{m+1,n} \, , \qquad
    R(c_{m,n}) := c_{m,n+1} \, ,
\ee
and 
\be
      L(a b) = L(a) \, b \, , \qquad 
      R(a b) = a \, R(b)  \;. 
\ee
Obviously, we have
\be
    c_{m,n} = L^m R^n(c) \, , \qquad  c := c_{0,0} \; . 
\ee
The \emph{free WNA algebra} $\A_{\mathrm{free}}(f)$ over $\mathcal{R}$ is 
then defined as the algebra $\mathcal{A}_{\mathrm{free}}$ augmented 
with an element $f$, such that 
\be
    f f = c \, , \qquad 
    f a = L(a) \, , \qquad 
    a f = R(a) \; .
\ee
It is easily seen that $f \not\in \A_{\mathrm{free}}(f)'$, thus 
$\A_{\mathrm{free}}(f)' = \mathcal{A}_{\mathrm{free}}$, 
and $f$ generates $\A_{\mathrm{free}}(f)$. 
Any other WNA algebra $\A(f')$ over $\mathcal{R}$, with a single generator, 
is the homomorphic image of $\A_{\mathrm{free}}(f)$ by the map given by $f \mapsto f'$ and 
$c_{m,n} \mapsto L_{f'}^m R_{f'}^n(f'{}^2)$ (cf. proposition~\ref{prop:span}). 
\vskip.2cm

In $\A_{\mathrm{free}}$ we have 
\be
    h_{n+1} = L^n(c) \, , \qquad e_{n+1} = R^n(c) \qquad \quad n=0,1,\ldots \; .
\ee

\begin{proposition}
\label{prop:number_of_hm}
For $n=0,1,2,\ldots$, the number of linearly independent monomials in $\A_{\mathrm{free}}(f)$, 
which are homogeneous of degree $n+2$ in $f$, is $2^n$.
\end{proposition}
{\it Proof:} 
The number of monomials $c_{r,s} = L_f^r R_f^s(f^2)$, $r,s \geq 0$, with fixed $r+s$ is 
$k := r+s+1$. Any monomial in $\A_{\mathrm{free}}(f)' = \mathcal{A}_{\mathrm{free}}$ 
containing $n+2$ $f$'s can be written in a unique way as a product 
$c_{r_1,s_1} \cdots c_{r_{j+1},s_{j+1}}$ where 
$n+2 = (r_1+s_1+2) + \cdots + (r_{j+1}+s_{j+1}+2)$ with some 
$j \in \{ 0,1,\ldots, \lfloor n/2 \rfloor \}$ (where $\lfloor n/2 \rfloor$ is 
the largest integer $\leq n/2$). 
For fixed $j$, the number of independent monomials is 
\bez
    \sum_{k_1+\cdots + k_{j+1} = n+1-j} k_1 k_2 \cdots k_{j+1}
  = {n+1 \choose 2j+1}  \; .
\eez
This combinatorial identity is obtained from
\bez
      \sum_{m \geq 0} {m+2j+1 \choose 2j+1} t^m 
  &=& \sum_{m \geq 0} {-2(j+1) \choose m} (-t)^m 
  = (1-t)^{-2(j+1)}  \\
  &=& \sum_{m \geq 0} \Big( \sum_{k_1+ \ldots+k_{j+1} = m+j+1} k_1 \cdots k_{j+1} \Big) \, t^m \, , 
\eez
where the last equality used $(1-t)^{-2} = \sum_{k \geq 1} k \, t^{k-1}$, which in turn 
results from $(1-t)^{-1} = \sum_{k \geq 0} t^k$ by differentiation.  
Our assertion now follows from
\bez
    \sum_{j=0}^{\lfloor n/2 \rfloor} {n+1 \choose 2j+1} = 2^n \; . 
\eez
\hfill $\square$

\section{A sequence of products}
\label{section:products}
\setcounter{equation}{0}
Let $\A$ be any (nonassociative) algebra and $f \in \A$. 
Then a sequence of products $\circ_n$, $n=1,2,\ldots$, in $\A$ 
is determined by $a \circ_1 b := a \, b$ and 
\be
    a \circ_{n+1} b
 := a \, (f \circ_n b) - (a f) \circ_n b  \qquad \quad 
         n = 1,2, \ldots \; .
    \label{circ_n}
\ee
These products provide us with a certain measure of nonassociativity at 
different levels. For example, 
\be
       a \circ_2 b
 &=& a \, (f b) - (a f) \, b = - (a,f,b) \, , \\
       a \circ_3 b
 &=& a \, (f (f b)) - a \, (f^2 \, b)
     - (a f)(f b) + ((a f)f) \, b  \nonumber \\
 &=& - a \, (f,f,b) + (a f,f,b)
\ee
If $f \in \A'$, then $a \circ_n b = 0$ for $n>1$. Hence we should take 
$f \in \A \setminus \A'$ in the following. 
For $a,b \in \A$, let us set 
\be
    L^{(m)}_a (b) := a \circ_m b \, , \qquad
    R^{(m)}_b (a) := a \circ_m b \; .
\ee
Moreover, we introduce
\be
    e_{(n_1,\ldots,n_r)} := R^{(n_r)}_f \cdots R^{(n_1)}_f (f) \, , \qquad 
    h_{(n_1,\ldots,n_r)} := L^{(n_1)}_f \cdots L^{(n_r)}_f(f) \, ,
       \label{eh_(...)}
\ee
for $r, n_1,\ldots,n_r=1,2,\ldots$. Note that
\be
    e_{(n)} = h_{(n)} = p_n  \, , \qquad
    e_{(1^n)} := e_{(1,1,\ldots,1)} = R_f^n(f) = e_n \, , \qquad
    h_{(1^n)} = h_n \qquad n=1,2, \ldots \, ,
\ee
where
\be
     p_n := f \circ_n f  \qquad \quad  n=1,2, \ldots \; .   \label{p_n}
\ee
\vskip.2cm

\noindent
{\bf Example.} In $\A_{\mathrm{free}}(f)$ (see section~\ref{subsection:freeWNA}), 
we have $p_1 = c$, and 
\be
   && p_2 = L(c) - R(c) \, , \qquad
      p_3 = L^2(c) - L R(c) + R^2(c) - c^2 \, ,  \nonumber \\
   && p_4 = L^3(c) - L^2R(c) + LR^2(c) - R^3(c) + c \, ( R(c) - L(c)) + (R(c)-L(c)) \, c \; .
\ee
\hfill $\square$
\vskip.2cm

In the following we derive some properties of the new products in the case where 
$\A$ is a WNA algebra. In particular, it turns out that the products $\circ_n$ 
then only depend on the equivalence class $[f] \in \A/\A'$ determined by $f$.

\begin{proposition}
\label{prop:ass_mn}
Let $\A$ be a WNA algebra. For all $m,n \in \mathbb{N}$, 
and all $a,c \in \A$, 
\be
    (a \circ_n b) \circ_m c
  = a \circ_n (b \circ_m c)  \qquad
    \mbox{if} \quad  b \in \A' \, ,     \label{prodn_assoc}
\ee
which can also be expressed as
\be
    [L_a^{(n)} , R_c^{(m)}] = 0 \qquad \mbox{on} \quad \A'  \; .  \label{gLRcommA'}
\ee
\end{proposition}
{\it Proof:}
We first prove the case $n=1$. Our assertion obviously holds for $m=1$. 
Assuming that it holds for $m$, the induction step is 
\bez
     (a b) \circ_{m+1} c
 &=& (a b) (f \circ_m c) - (a b f) \circ_m c
  = a \, ( b \, (f \circ_m c))
     - a \, ((b f) \circ_m c)   \\
 &=& a \, ( b \, (f \circ_m c)
     - (b f) \circ_m c )
  = a \, (b \circ_{m+1} c)
\eez
for $b \in \A'$. 
Now we prove (\ref{prodn_assoc}) by induction on $n$. The corresponding 
induction step is
\bez
     a \circ_{n+1} (b \circ_m c)
 &=& a \, (f \circ_n (b \circ_m c))
     - (a f) \circ_n (b \circ_m c)  \\
 &=& a \, ((f \circ_n b) \circ_m c)
     - ((a f) \circ_n b) \circ_m c
     \qquad \mbox{by induction hypothesis}   \\
 &=& (a \, (f \circ_n b)) \circ_m c
     - ((a f) \circ_n b) \circ_m c
      \qquad \mbox{since }  f\circ_n b \in \A'  \\
 &=& (a \circ_{n+1} b) \circ_m c) \; .
\eez
\hfill $\square$

\begin{lemma}
\label{lemma:ass_alt}
Let $\A$ be a WNA algebra. For all $n \in \mathbb{N}$,
\be
    a \circ_{n+1} b = a \circ_n (f b)
     - (a \circ_n f) \, b \; .   \label{prodn_alt}
\ee
\end{lemma}
{\it Proof:}
By definition this holds for $n=1$ and it is easily verified for $n=2$. 
Let us assume that it holds for $n+1$. Then we have 
\bez
      a \circ_{n+2} b
  &=& a \, (f \circ_{n+1}b) - (a f) \circ_{n+1} b  \\
  &=& a \, (f \circ_n (f b)) - a \, (f \circ_n f) \, b
      - (a f) \circ_n (f b) + ((a f) \circ_n f) b
\eez
by use of the induction hypothesis and $f \circ_n f \in \A'$. 
Combining the first term with the third and the second with the fourth, 
we obtain
\bez
     a \circ_{n+2} b
 &=& a \circ_{n+1} (f b) - (a \circ_{n+1} f) \, b \; .
\eez
\hfill $\square$
\bigskip

More generally, we have the following result.

\begin{proposition}
\label{prop:circ_m_circ_n-ass}
Let $\A$ be a WNA algebra. For all $m,n \in \mathbb{N}$,
\be
   a \circ_m (f \circ_n b) - (a \circ_m f) \circ_n b
   = a \circ_{m+n} b \, ,   \label{circ_m_circ_n-ass}
\ee
which can also be expressed as
\be
    [L^{(m)}_a , R^{(n)}_b](f) = L^{(m+n)}_a(b) = R^{(m+n)}_b(a) \; . \label{gLRcommf}
\ee
\end{proposition}
{\it Proof:}
By definition this holds for $m=1$ and all $n$. 
Let us assume that it holds for $m$ (and all $n$). Using (\ref{prodn_alt}) 
and (\ref{prodn_assoc}), we find
\bez
  & & a \circ_{m+1} (f \circ_n b) - (a \circ_{m+1} f) \circ_n b \\
  &=& a \circ_m (f(f \circ_n b)) - (a \circ_m f)(f \circ_n b)
     - (a \circ_m f^2) \circ_n b + ((a \circ_m f)f) \circ_n b \\
  &=& a \circ_m (f(f \circ_n b)) - (a \circ_m f)(f \circ_n b)
     - a \circ_m (f^2 \circ_n b) + ((a \circ_m f)f) \circ_n b
     \; .
\eez
Grouping the first with the third term and the second with the fourth, we obtain 
\bez
    a \circ_{m+1} (f \circ_n b) - (a \circ_{m+1} f) \circ_n b
  = a \circ_m (f \circ_{n+1} b) - (a \circ_m f) \circ_{n+1} b
\eez
which, by use of the induction hypothesis, proves our assertion.
\hfill $\square$

\begin{proposition}
\label{prop:circ_n-equiv}
Let $\A$ be a WNA algebra. Then the products $\circ_n$ only depend on the 
equivalence class $[f] \in \A/\A'$.
\end{proposition}
{\it Proof:} For all $b \in \A'$, we have
\bez
    a \, ((f + b) \circ_n c) - (a (f + b)) \circ_n c
 =  a \circ_{n+1} c
    + a \, ( b \circ_n c) - (a b) \circ_n c
 = a \circ_{n+1} c
\eez
by use of proposition~\ref{prop:ass_mn}.
\hfill $\square$

\begin{lemma}
Let $\A$ be a WNA algebra. Then the following relations hold.
\be
       f \circ_k e_{(n_1,\ldots,n_r)} 
 &=& e_{(k,n_1,\ldots,n_r)} + e_{(k+n_1,n_2,\ldots,n_r)} \, ,   \label{f-circ-e_}  \\
        e_{(n_1,\ldots,n_r)} \circ_k e_{(m_1,\ldots,m_s)} 
 &=& e_{(n_1,\ldots,n_r,k,m_1,\ldots,m_s)} + e_{(n_1,\ldots,n_r,k+m_1,\ldots,m_s)} \; . 
\ee
\end{lemma} 
{\it Proof:} The first relation is obtained as follows,
\bez
      L_f^{(k)} e_{(n_1,\ldots,n_r)} 
  &=& R^{(n_r)}_f \cdots R^{(n_2)}_f L_f^{(k)} R^{(n_1)}_f(f)  \\
  &=& R^{(n_r)}_f \cdots R^{(n_2)}_f ( R^{(n_1)}_f L_f^{(k)}(f) + R^{(k+n_1)}_f(f) ) \\
  &=& e_{(k,n_1,\ldots,n_r)} + e_{(k+n_1,n_2,\ldots,n_r)} \, ,
\eez
where we applied (\ref{gLRcommA'}) and (\ref{gLRcommf}). 
Furthermore, using (\ref{circ_m_circ_n-ass}), we have
\bez
     e_{(n_1,\ldots,n_r)} \circ_k (f \circ_m f) 
 &=& (e_{(n_1,\ldots,n_r)} \circ_k f) \circ_m f 
     + e_{(n_1,\ldots,n_r)} \circ_{k+m} f \\
 &=& e_{(n_1,\ldots,n_r,k,m)} + e_{(n_1,\ldots,n_r,k+m)} \, ,
\eez
and then also 
\bez
     e_{(n_1,\ldots,n_r)} \circ_k e_{(m_1,\ldots,m_s)}
 &=& e_{(n_1,\ldots,n_r)} \circ_k R_f^{(m_s)} \cdots R_f^{(m_2)}(f \circ_{m_1} f) \\
 &=& R^{(m_s)}_f \cdots R^{(m_2)}_f(e_{(n_1,\ldots,n_r)} \circ_k (f\circ_{m_1} f)) \\
 &=& R^{(m_s)}_f \cdots R^{(m_2)}_f(e_{(n_1,\ldots,n_r,k,m_1)} + e_{(n_1,\ldots,n_r,k+m_1)}) \\
 &=& e_{(n_1,\ldots,n_r,k,m_1,\ldots,m_s)} + e_{(n_1,\ldots,n_r,k+m_1,m_2,\ldots,m_s)} \, ,
\eez
by use of (\ref{prodn_assoc}). 
\hfill$\square$

\begin{proposition}
A WNA algebra $\A(f)$ is spanned by $f$ and $e_{(n_1,\ldots,n_r)}$, where 
$r,n_1,\ldots,n_r=1,2,\ldots$.
\end{proposition}
{\it Proof:} According to proposition (\ref{prop:span}), the preceding lemma, 
and $e_{(n_1,\ldots,n_r)} \circ_k f = e_{(n_1,\ldots,n_r,k)}$ (which holds by definition), 
it is sufficient to show that elements of the form $L_f^n R_f^m(f^2)$ can be expressed 
as linear combinations of elements $e_{(k_1,\ldots,k_r)}$. But this follows from
\bez
    L_f^n R_f^m(f^2) = L_f^n R_f^m(e_{(1)}) = L_f^n(e_{(1^{m+1})})
\eez
and iterated application of (\ref{f-circ-e_}).
\hfill$\square$
\bigskip

Obviously, $e_{(k_1,\ldots,k_r)}$ is homogeneous  in $f$ of degree $k_1+\cdots+k_r+1$. 
The number of monomials $e_{(k_1,\ldots,k_r)}$ in $\A_{\mathrm{free}}(f)$, 
which are homogeneous in $f$ of degree $n+2$ is therefore equal to the number of compositions 
$(k_1,\ldots,k_r)$ of $n+1$, which is $2^n$, in accordance with 
proposition~\ref{prop:number_of_hm}.
The last lemma and proposition can also be formulated in terms of the elements 
$h_{(n_1,\ldots,n_r)}$, of course.

\section{Derivations of WNA algebras}
\label{section:derivations}
\setcounter{equation}{0}
In this section we construct a sequence of commuting derivations for special 
WNA algebras. First we note a general property of derivations of WNA algebras.

\begin{proposition}
\label{prop:deriv-circ_n}
Any derivation $\delta$ of a WNA algebra $\A$ with the property $\delta(\A) \subset \A'$ 
is also a derivation with respect to any of the products $\circ_n$, $n \in \mathbb{N}$.
\end{proposition}
{\it Proof:} By induction. The induction step can be formulated as 
follows,
\bez
     \delta(a \circ_{n+1} b)
 &=& \delta( a \, (f \circ_n b) - (a f) \circ_n b )  \\
 &=& \delta(a) \, (f \circ_n b)
     + a \, (\delta(f) \circ_n b)
     + a \, (f \circ_n \delta(b))  \\
 & & - (\delta(a) \, f) \circ_n b
     - (a \, \delta(f)) \circ_n b
     - (a \, f) \circ_n \delta(b) \\
 &=& \delta(a) \, (f \circ_n b)
     + a \, (f \circ_n \delta(b))
     - (\delta(a) \, f) \circ_n b
     - (a f) \circ_n \delta(b)  \\
 &=& \delta(a) \circ_{n+1} b + a \circ_{n+1} \delta(b) \, ,
\eez
where we used (\ref{prodn_assoc}). 
\hfill $\square$
\bigskip

We call a subalgebra $\A(f)$ of a nonassociative WNA algebra $\A$ 
\emph{$\delta$-compatible}, if it admits derivations 
$\delta_n$, $n = 1,2, \ldots$, such that
\be
    \delta_n(f) := p_n \equiv f \circ_n f   \qquad \quad n = 1,2, \ldots \; .
                \label{delta_n-def}
\ee
 For $n=1,2,3$, the definition (\ref{delta_n-def}) reproduces (\ref{hier1}), 
(\ref{hier2}) with $\kappa_1=\kappa_2=1$, and (\ref{hier3}), respectively. 
Clearly, $\A_{\mathrm{free}}(f)$ is $\delta$-compatible. If 
$\mathcal{I}$ is a two-sided ideal in $\A_{\mathrm{free}}(f)$, which is invariant 
under the action of the derivations $\delta_n$, i.e., 
$\delta_n(\mathcal{I}) \subset \mathcal{I}$, $n=1,2,\ldots$, 
then $\A_{\mathrm{free}}(f)/\mathcal{I}$ is also $\delta$-compatible, since the 
derivations of $\A_{\mathrm{free}}(f)$ project to derivations of 
$\A_{\mathrm{free}}(f)/\mathcal{I}$. 
\vskip.2cm

\noindent
{\bf Example.} 
Let $\mathcal{I}$ be the two-sided ideal in $\A_{\mathrm{free}}(f)$ generated by 
a subset of the elements $p_n$, $n=1,2,\ldots$. 
As a consequence of the next proposition, 
$\delta_m(p_k) = \delta_k(p_m) = p_k \circ_m f + f \circ_m p_k$, so that 
$\mathcal{I}$ is invariant and $\A_{\mathrm{free}}(f)/\mathcal{I}$ 
is $\delta$-compatible. Of course, if $p_k \in \mathcal{I}$, then $\delta_k \equiv 0$ 
on $\A_{\mathrm{free}}(f)/\mathcal{I}$. 
\hfill $\square$
\vskip.2cm

In the following we assume that $\A(f)$ is a $\delta$-compatible 
WNA algebra.

\begin{proposition}
\label{prop:comm} 
The derivations $\delta_n$, $n=1,2,\ldots$, commute with each other on $\A(f)$.
\end{proposition}
{\it Proof:} Proposition (\ref{prop:circ_m_circ_n-ass}) implies
\be
   f \circ_m p_n - p_m \circ_n f = p_{m+n} = f \circ_n p_m - p_n \circ_m f \; .
   \label{p_m+n}
\ee
Hence
\bez
     \delta_m \delta_n(f)
 &=& \delta_m(p_n)
  =  \delta_m(f \circ_n f)
  = \delta_m(f) \circ_n f + f \circ_n \delta_m(f) \\
 &=& p_m \circ_n f + f \circ_n p_m
  = p_n \circ_m f + f \circ_m p_n   \\
 &=& \delta_n(f) \circ_m f + f \circ_m \delta_n(f)
  = \delta_n(p_m)   \\
 &=& \delta_n \delta_m(f) \; .
\eez
\hfill $\square$
\vskip.2cm

Next we determine the action of the derivations on the monomials $e_{(m_1,\ldots,m_r)}$ 
defined in (\ref{eh_(...)}).

\begin{lemma}
\be
    [\delta_n , R^{(m)}_f] = R^{(n)}_f R^{(m)}_f + R^{(n+m)}_f   
        \qquad \quad \mbox{on} \quad \A(f) \; .
\ee
\end{lemma}
{\it Proof:} For $a \in \A(f)$, we have
\bez
    [\delta_n , R^{(m)}_f](a) 
  = a \circ_m (f \circ_n f) = (a \circ_m f) \circ_n f + a \circ_{n+m} f \, ,
\eez
by use of (\ref{circ_m_circ_n-ass}).
\hfill$\square$

\begin{proposition}
\label{prop:delta(e_)}
\be
    \delta_n(e_{(m_1,\ldots,m_r)}) 
  = f \circ_n e_{(m_1,\ldots,m_r)} 
    + \sum_{k=1}^{r-1}e_{(m_1,\ldots,m_k)} \circ_n e_{(m_{k+1},\ldots,m_r)}
    + e_{(m_1,\ldots,m_r)} \circ_n f \; .    \label{delta(e_(...))}
\ee
\end{proposition}
{\it Proof:} We use induction on $r$. For $r=1$, we have
\bez
    \delta_n (e_{(m_1)}) 
  = \delta_n (p_{m_1}) 
  = \delta_{m_1}(f \circ_n f)
  = (\delta_{m_1}(f)) \circ_n f + f \circ_n (\delta_{m_1}(f))
  = e_{(m_1)} \circ_n f + f \circ_n e_{(m_1)} \, ,
\eez
where the second step made use of proposition~\ref{prop:comm}. 
Let us assume that the formula holds for some $r \geq 1$. 
With the help of the preceding lemma, we find
\bez
     \delta_n(e_{(m_1,\ldots,m_{r+1})})
 &=& \delta_n R^{(m_{r+1})}_f(e_{(m_1,\ldots,m_r)}) \\
 &=& e_{(m_1,\ldots,m_{r+1})} \circ_n f + e_{(m_1,\ldots,m_{r+1}+n)}
    + R^{(m_{r+1})}_f \delta_n (e_{(m_1,\ldots,m_r)}) \; .
\eez
Using the induction hypothesis and (\ref{prodn_assoc}), the last term 
can be expressed as follows,
\bez
     R^{(m_{r+1})}_f \delta_n (e_{(m_1,\ldots,m_r)})
 &=& f \circ_n e_{(m_1,\ldots,m_{r+1})}
     + \sum_{k=1}^{r-1} e_{(m_1,\ldots,m_k)} \circ_n e_{(m_{k+1},\ldots,m_{r+1})} \\
 & & + (e_{(m_1,\ldots,m_r)} \circ_n f) \circ_{m_{r+1}} f \; .
\eez
Now we have to use (\ref{circ_m_circ_n-ass}) to see that our assertion holds for $r+1$.
\hfill$\square$

\section{KP identities}
\label{section:KPids}
\setcounter{equation}{0}
In this section we consider a $\delta$-compatible subalgebra $\A(f)$ of a 
WNA algebra $\A$. We derive identities for the elements 
$\delta_{n_1} \cdots \delta_{n_r}(f)$ and establish a correspondence with 
equations of the potential KP hierarchy.

\begin{lemma}
\be
    p_n &=& h_n - \sum_{k=1}^{n-1} p_k \, h_{n-k-1}   \label{p_n=h_n...} \\
    p_n &=& (-1)^{n+1} e_n - \sum_{k=1}^{n-1} (-1)^{n-k} \, e_{n-k-1}\,p_k
             \; .   \label{p_n=e_n...}
\ee
\end{lemma}
{\it Proof:}
With the help of proposition~\ref{prop:circ_m_circ_n-ass}, we obtain
\bez
       \sum_{k=1}^{n-1} p_k \, h_{n-k-1}
 &=& \sum_{k=1}^{n-1} \Big( f \circ_k (f \, h_{n-k-1})
               - f \circ_{k+1} h_{n-k-1} \Big)           \\
 &=& \sum_{k=1}^{n-1} (f \circ_k h_{n-k} - f \circ_{k+1} h_{n-k-1})
  = f \circ_1 h_{n-1} - f \circ_n h_0
  = h_n - p_n \; .
\eez
The second formula is proved in a similar way, using 
$e_{n+1} \circ_m f + e_n \circ_{m+1} f = e_n \, p_m$.
\hfill $\square$

\begin{lemma}
\label{lemma:he_n=sumdelta}
\be
    n \, h_n = \sum_{k=1}^n \delta_k(h_{n-k}) \, , \qquad
    n \, e_n = \sum_{k=1}^n (-1)^{k+1} \delta_k(e_{n-k})
    \qquad n=1,2, \ldots \; .     \label{he_n=sumdelta}
\ee
\end{lemma}
{\it Proof:} For $n=1$ this follows from the definitions. 
Let us assume that it holds for $n-1$. Using the derivation rule and the 
definitions, we find
\bez
    p_k h_{n-k-1} = \delta_k(h_{n-k}) - f \, \delta_k(h_{n-k-1}) \; .
\eez
With its help and by use of (\ref{p_n=h_n...}), we obtain
\bez
      h_n - p_n
 &=& \sum_{k=1}^{n-1} p_k h_{n-k-1}
  = \sum_{k=1}^n \delta_k(h_{n-k}) - \delta_n(h_0)
    - f \sum_{k=1}^{n-1} \delta_k(h_{n-1-k})
\eez
and, using the induction hypothesis,
\bez
   h_n = \sum_{k=1}^n \delta_k(h_{n-k}) - (n-1) f \, h_{n-1}
\eez
so that
\bez
   n \, h_n = \sum_{k=1}^n \delta_k(h_{n-k}) \; .
\eez
The second formula is proved analogously.
\hfill $\square$
\bigskip

\begin{theorem}
\label{theorem:Schur}
\be
    h_n = \chi_n(\tilde{\delta})(f) \, , \qquad
    e_n = (-1)^n \chi_n(-\tilde{\delta})(f)
\ee
where $\chi_n$ are the elementary Schur polynomials defined by
\be
     \exp( \sum_{k \geq 1} \lambda^k \, t_k )
   = \sum_{n \geq 0} \chi_n(t_1,t_2,\ldots) \, \lambda^n
         \label{Schur} 
\ee
with independent variables $t_k$, $k=1,2, \ldots$, an indeterminate $\la$, and
\be
  \tilde{\delta} := (\delta_1, \delta_2/2, \delta_3/3, \ldots)  \; .
  \label{tildedelta}
\ee
\end{theorem}
{\it Proof:} In terms of the formal power series
\bez
    h(\la) := \sum_{n \geq 0} \la^n h_n \, , \qquad
    e(\la) := \sum_{n \geq 0} \la^n e_n \, , \qquad
    \delta_\la := \sum_{n \geq 1} \la^{n-1} \delta_n \, , 
\eez
equations (\ref{he_n=sumdelta}) read
\bez
    {d \over d \la} h(\la) = \delta_\la (h(\la)) \, , \qquad
    {d \over d \la} e(-\la) = - \delta_\la (e(-\la)) \; .
\eez
Integration leads to
\bez
  h(\la) = \exp \left(\sum_{n \geq 1} {\la^n \over n} \delta_n \right) f \, , \qquad
  e(\la) = \exp \left(-\sum_{n \geq 1} {(-\la)^n \over n} \delta_n \right) f
\eez
where the constant of integration is fixed by $h(0)=f=e(0)$. Our assertions 
are now verified by comparison with the generating formula (\ref{Schur}) 
for the elementary Schur polynomials.
\hfill $\square$
\bigskip

Introducing the abbreviations
\be
    H_n := \chi_n(\tilde{\delta}) \, , \qquad
    E_n := (-1)^n \chi_n(-\tilde{\delta}) \, ,  \label{HnEn-def}
\ee
where $H_0 = \mathrm{id} = E_0$, theorem~\ref{theorem:Schur} states that
\be
    h_n = H_n(f) \, , \qquad
    e_n = E_n(f) \; .          \label{HnEn-basic}
\ee
All $H_n, E_n$ mutually commute as a consequence of proposition~\ref{prop:comm}. 
 From (\ref{HnEn-basic}) and the definition of $h_n$ and $e_n$, 
we obtain
\be
    H_{n+1}(f) = f \, H_n(f) 
    \, , \qquad
    E_{n+1}(f) = E_n(f) \, f  \; .    \label{HE-recur}
\ee
\vskip.2cm

Since, according to propositions~\ref{prop:deriv-circ_n} and \ref{prop:comm}, 
the $\delta_n$ are commuting derivations of $\A(f)$ with respect to all the products 
introduced in section~\ref{section:products}, the formal power series
\be
  H(\la) &:=& \sum_{n \geq 0} \la^n H_n
           = \exp \left(\sum_{n \geq 1} {\la^n \over n} \delta_n \right)
             \, , \\
  E(\la) &:=& \sum_{n \geq 0} \la^n E_n
           = \exp \left(-\sum_{n \geq 1} {(-\la)^n \over n} \delta_n \right)
             \, ,         \label{H(la)E(la)-def}
\ee
are homomorphisms of all these products. Here $\mathcal{R}$ has to be extended to 
the ring $\mathcal{R}[[\la]]$ of formal power series in $\la$. Note that
\be 
    E(-\la) \, H(\la) = \mathrm{id}  \; .   \label{inveh}
\ee
\vskip.2cm

\begin{proposition}
 For $n=0,1,2, \ldots$, $m = 1,2, \ldots$, and for all
 $a, b \in \A(f)$,
\be
      H_n( a \circ_m b )
  &=& \sum_{k=0}^n H_k(a) \circ_m H_{n-k}(b)
                   \label{H_gen_derivation} \\
      E_n( a \circ_m b )
  &=& \sum_{k=0}^n E_k(a) \circ_m E_{n-k}(b) \; .
                   \label{E_gen_derivation}
\ee
Thus $\{ H_n \}_{n=0}^\infty$ and $\{ E_n \}_{n=0}^\infty$ 
are Hasse-Schmidt derivations \cite{Hass+Schm37,Mats86}.
\end{proposition}
{\it Proof:}
This follows by expansion of
\bez
  H(\la)(a \circ_m b) = H(\la)(a) \circ_m H(\la)(b) \, , \qquad
  E(\la)(a \circ_m b) = E(\la)(a) \circ_m E(\la)(b)
\eez
in powers of $\la$.
\hfill $\square$

\begin{theorem}
\label{theorem:algebraicFay}
\be
    - \delta_1 \Big( (H_+(\la_1) - H_+(\la_2))(f) \Big)
  &=& \Big( {1 \over \la_1} - {1 \over \la_2}
      + (H_+(\la_1) - H_+(\la_2))(f) \Big) \; H_+(\la_1) \, H_+(\la_2)(f)
      \nonumber \\
  &&  + [H_+(\la_1)(f) , H_+(\la_2)(f)]  \; .  \label{algebraicFay}
\ee
where $H_+(\la) := H(\la) - H_0$. 
\end{theorem}
{\it Proof:} The first of equations (\ref{HE-recur}) can be written as
\bez
    H(\la)(f) = f + \la f \, H(\la)(f)  \; .
\eez
In terms of $H_+(\la)$, this takes the form
\bez
  0 = {1 \over \la} H_+(\la)(f) - f^2 - f \, H_+(\la)(f)
    = {1 \over \la} H_+(\la)(f) - \delta_1(f) - f \, H_+(\la)(f)
\eez
where we used $\delta_1(f) = f^2$. 
Replacing $\la$ by $\la_1$, and acting with $H_+(\la_2)$ on this equation, 
we obtain
\bez
  && {1 \over \la_1} H_+(\la_2) \, H_+(\la_1)(f) - H_+(\la_2)(\delta_1(f))
     \nonumber \\
 &=& H_+(\la_2)(f) \, H_+(\la_2)H_+(\la_1)(f)
   + H_+(\la_2)(f) \, H_+(\la_1)(f)
     + f \, H_+(\la_2) \, H_+(\la_1)(f)  \; .
\eez
Antisymmetrization in $\la_1,\la_2$, eliminates terms involving a `bare' $f$ 
(i.e., without a $\delta_n$ acting on it) and leads to (\ref{algebraicFay}).
\hfill $\square$
\bigskip

Expanding (\ref{algebraicFay}) in powers of $\la_1, \la_2$, yields, 
for $m,n =1,2,\ldots$, 
\be
   H_m H_{n+1}(f) - H_n H_{m+1}(f)
 = \sum_{k=1}^m H_k(f) \, H_{m-k} H_n(f)
   - \sum_{k=1}^n H_k(f) \, H_{n-k} H_n(f) \; .   \label{H-KPmn}
\ee

\begin{corollary}
\label{cor:functionalKP-id}
\be
    \sum_{i,j,k=1}^3  \varepsilon_{ijk} \, 
       \Big( \la_i^{-1} - H_+(\la_k)(f) \Big ) \, H(\la_k) \, H_+(\la_i)(f) = 0  
            \, ,  \label{functionalKP-id}
\ee
where $\varepsilon_{ijk}$ is totally antisymmetric with $\varepsilon_{123} = 1$.
\end{corollary}
{\it Proof:}  This follows by adding (\ref{algebraicFay}) three times with 
cyclically permuted indeterminates $\la_1, \la_2, \la_3$.
Alternatively, we can start from 
\bez
   \sum_{i,j,k=1}^3 \varepsilon_{ijk} \, H(\la_k) \Big( \la_i^{-1} (H(\la_i)(f)-f)
       - f \, H(\la_i)(f) \Big) = 0 
\eez
which is a trivial consequence of (\ref{HE-recur}) (in the form of the first 
equation in the proof of theorem~\ref{theorem:algebraicFay}). This is easily 
shown to be equivalent to (\ref{functionalKP-id}). 
\hfill $\square$
\bigskip

It is important to note that all factors appearing in (\ref{algebraicFay}) and 
(\ref{functionalKP-id}) lie in the associative subalgebra $\A(f)'$. 
The first non-trivial identity which results from expanding these functional equations 
in powers of the indeterminates, is (\ref{KP-id}), which has the form 
of the potential KP equation. 
In fact, if we replace $\tilde{\delta}$ by
\be
      \tilde{\pa} := (\pa_{t_1}, \pa_{t_2}/2, \pa_{t_3}/3, \ldots)
\ee
with partial derivatives $\pa_{t_n}$ with respect to independent variables 
$t_n$, $n=1,2, \ldots$, then equation (\ref{functionalKP-id}) becomes a 
generating formula for the potential KP hierarchy as derived in 
\cite{Bogd+Kono98,Bogd99} (see also section~\ref{subsection:KP}). 
This proves an assertion formulated in the introduction. 
Of course, only with the choice $\A_{\mathrm{free}}(f)$ we obtain the full 
set of KP hierarchy equations in this way. Other choices for the $\delta$-compatible 
WNA algebra lead to reductions of the KP hierarchy. 
The KP hierarchy makes its appearance in many areas of mathematics (in particular 
differential and algebraic geometry) and physics (from hydrodynamics to string theory), 
and the above result further adds to its ubiquitousness. 
\vskip.2cm

There are, of course, analogs of (\ref{algebraicFay}) and (\ref{functionalKP-id}) 
formulated in terms of $E(\la)$ instead of $H(\la)$. The analog of (\ref{H-KPmn}) is
\be
   E_m E_{n+1}(f) - E_n E_{m+1}(f)
 = \sum_{k=1}^m E_{m-k} E_n(f) \, E_k(f)
    - \sum_{k=1}^n E_{n-k} E_m(f) \, E_k(f)    \label{E-KPmn}
\ee
for $m,n =1,2,\ldots$.
The KP hierarchy in this form appeared in \cite{DNS89}, for example.

\section{Quasi-symmetric functions and KP identities}
\label{section:qs}
\setcounter{equation}{0}
Some results in the preceding section should remind us of formulae for  
symmetric functions. In this section we construct a WNA algebra 
that contains an element $f$ which generates the algebra of 
\emph{quasi-symmetric functions} 
(see \cite{Gess84,Reut93,Malv+Reut95,Haze00}, for example). 
The main result of section~\ref{section:KPids} can then be applied: 
the `KP identities' given by theorem~\ref{theorem:algebraicFay} (or 
corollary~\ref{cor:functionalKP-id}) determine corresponding identities in 
the algebra of quasi-symmetric functions.
\vskip.2cm

Let $\mathcal{A} = \mathbb{Z}[[x_1,x_2,\ldots]]$ be the ring (algebra over 
$\mathcal{R} = \mathbb{Z}$) of formal power series in a set of commuting 
indeterminates $x_n$, $n=1,2,\dots$, with product denoted by concatenation 
(which should not be confused with our previous notation for the product 
in a WNA algebra, for which we will only use $\circ_1$ below). 
A monomial $a$ in $\mathcal{A}$ is of the form $a = x_{i_1} \cdots x_{i_r}$, 
and we set
\be
    m(a) := \min\{ i_1,\ldots,i_r \} \, , \qquad     
    M(a) := \max\{ i_1,\ldots,i_r \} \; .
\ee
Another product in $\mathcal{A}$ is then given by
\be
       a \circ_1 b
   := a \, b \, \sum_{M(a) < i \leq m(b)}  x_i \qquad \mbox{if} \quad  M(a) < m(b) \, , 
\ee
and $a \circ_1 b =0$ if $M(a) \geq m(b)$, where $a,b$ are any monomials. 
Then $(\mathcal{A}, \circ_1)$ is an associative algebra. 
\vskip.2cm

Next we augment $(\mathcal{A}, \circ_1)$ with an element $f$ and extend the 
product as follows, 
\be
  f \circ_1 f := \sum_{i} x_i  \, , \qquad
  f \circ_1 a := a \, \sum_{i \leq m(a)} x_i \, , \qquad
  a \circ_1 f := a \, \sum_{M(a) < i} x_i \; . \label{qs-prod2}
\ee
This determines a WNA algebra $(\A, \circ_1)$. Obviously, we have
\be
    e_n = \sum_{i_1 < \cdots < i_n} x_{i_1} \cdots x_{i_n}
           \, , \qquad
    h_n = \sum_{i_1 \leq \cdots \leq i_n} x_{i_1} \cdots x_{i_n}  \, ,
\ee
which are, respectively, the elementary and complete symmetric functions 
\cite{Macd95}.

\begin{proposition}
For $k,m,n = 1,2, \ldots$, we have
\be
    p_k \equiv f \circ_k f &=& \sum_i x_i^k \, ,  \label{qsf-p_n} \\
    f \circ_k a &=& a \, \sum_{i \leq m(a)} x_i^k \, ,  \label{qs-circ_k-2} \\
    a \circ_k f &=& a \, \sum_{M(a)<i} x_i^k  \, ,  \label{qs-circ_k-3} \\
     a \circ_k b 
 &=& a \, b \, \sum_{M(a) < i \leq m(b)} x_i^k  \qquad \mbox{if} \quad M(a) < m(b) \, ,  
                 \label{qs-circ_k-4}
\ee
and $a \circ_k b = 0$ if $M(a) \geq m(b)$. 
\end{proposition}
\textit{Proof:} By definition these relations hold for $k=1$. Let us assume 
that they hold for $k$. Then  
\bez
        f \circ_{k+1} f
    &=& f \circ_1 (f \circ_k f) - (f \circ_1 f) \circ_k f
     = \sum_i f \circ_1 x_i^k - \sum_i x_i \circ_k f  \\
    &=& \sum_{i \leq j} x_i \, x_j^k - \sum_{i<j} x_i \, x_j^k
     = \sum_i x_i^{k+1} 
\eez
shows that (\ref{qsf-p_n}) holds for $k+1$. The other relations are proved 
in a similar way. 
\hfill $\square$
\bigskip

Recalling the definitions (\ref{eh_(...)}), we obtain
\be
       e_{(n_1,\ldots,n_r)}
   = \sum_{i_1< \cdots <i_r} x_{i_1}^{n_1} \cdots x_{i_r}^{n_r} \, , \qquad
       h_{(n_1,\ldots,n_r)}
   = \sum_{i_1\leq \cdots \leq i_r} x_{i_1}^{n_1} \cdots x_{i_r}^{n_r} \; .
\ee
These sets of functions span the space of quasi-symmetric functions. 
Thus we arrive at the conclusion that the whole set of quasi-symmetric functions 
is generated by $f$ with the product $\circ_1$. In particular, the space of 
quasi-symmetric functions is closed under the product $\circ_1$. 
Since the expressions for the elements $e_{(n_1,\ldots,n_r)}$ are linearly 
independent, $\A(f)$ is in fact freely generated by $f$ (and thus isomorphic 
to $\A_{\mathrm{free}}(f)$). As a consequence, 
the construction of commuting derivations in section~\ref{section:derivations} 
and the results in section~\ref{section:KPids} apply in the case under consideration.

\begin{proposition}
\label{prop:qs_delta=p}
For $n=1,2, \ldots$, we have
\be
    \delta_n(a) = p_n \, a  \qquad \quad \forall a \in \A(f)'  \; .
\ee
\end{proposition}
{\it Proof:} 
Since $a \in \A(f)'$, so that $a$ is a quasi-symmetric function, 
it is sufficient to consider the case 
$a = e_{(m_1,\ldots,m_r)}$. The shuffle relation
\bez
     p_n \, e_{(m_1,\ldots,m_r)} 
 &=& \Big( \sum_i x_i^n \Big) \Big( \sum_{i_1< \cdots < i_r} x_{i_1}^{m_1} 
       \cdots x_{i_r}^{m_r} \Big) \\
 &=& \sum_{i \leq i_1 < \cdots < i_r} x_i^n x_{i_1}^{m_1} \cdots x_{i_r}^{m_r}
    + \sum_{i_1 <i \leq i_2 < \cdots < i_r} x_{i_1}^{m_1} x_i^n \, x_{i_2}^{m_2} 
        \cdots x_{i_r}^{m_r} + \cdots \\
 & & + \sum_{i_1 < \cdots < i_r <i} x_{i_1}^{m_1} \cdots x_{i_r}^{m_r} x_i^n 
\eez
can be expressed, with the help of (\ref{qs-circ_k-2})-(\ref{qs-circ_k-4}), as 
follows,
\bez
    p_n \, e_{(m_1,\ldots,m_r)}
  = f \circ_n e_{(m_1,\ldots,m_r)} 
    + \sum_{k=1}^{r-1} e_{(m_1,\ldots,m_k)} \circ_n e_{(m_{k+1},\ldots,m_r)}
    + e_{(m_1,\ldots,m_r)} \circ_n f \, , 
\eez
which, by use of proposition~\ref{prop:delta(e_)}, is 
\bez
    \delta_n(e_{(m_1,\ldots,m_r)}) = p_n \, e_{(m_1,\ldots,m_r)} \; .
\eez
\hfill $\square$
\bigskip

With the help of the last proposition and $\delta_n(f) = p_n$, the `KP identity' 
(\ref{KP-id}) leads to
\be
 & & \Big(\sum_i x_i \Big) \Big[ 4 \, \sum_i x_i^3 
     - \Big(\sum_i x_i \Big)^3 
     + 6 \, \sum_{i<j \leq k} x_i x_j x_k \Big] 
     - 3 \, \Big(\sum_i x_i^2 \Big)^2    \nonumber \\
 & & + 6 \, \sum_{i<j \leq k} \left( x_i^2 x_j x_k - x_i x_j x_k^2 \right) = 0  \; .
\ee
This is the first of the infinite sequence of `KP identities' in the algebra 
of quasi-symmetric functions, obtained from 
theorem~\ref{theorem:algebraicFay} (or corollary~\ref{cor:functionalKP-id}).
\vskip.2cm

Let $y_n$, $n=1,2,\ldots$, be a second set of commuting indeterminates, 
which commute with the members of the first set, and let $\mathcal{A}$ denote 
the ring $\mathbb{Z}[[x_1,x_2,\ldots,y_1,y_2,\ldots]]$. 
A monomial $a$ in $\mathcal{A}$ is now of the form 
$a = x_{i_1} \cdots x_{i_r} \, y_{j_1} \cdots y_{j_s}$, and we set
\be
    m(a) := \min\{ i_1,\ldots,i_r, j_1, \ldots, j_s \} \, , \qquad     
    M(a) := \max\{ i_1,\ldots,i_r, j_1, \ldots, j_s \} \; .
\ee
The above products $\circ_k$, $k=1,2,\ldots$, then generalize as follows,
\be
    f \circ_k f &=& \sum_i (x_i^k - y_i^k) \, ,  \\
          f \circ_k a 
  &=& a \, \Big( \sum_{i \leq m(a)} x_i^k - \sum_{i<m(a)} y_i^k \Big) \, ,  \\
          a \circ_k f
  &=& a \, \Big( \sum_{M(a)<i} x_i^k - \sum_{M(a) \leq i} y_i^k \Big) \, ,  \\
          a \circ_k b
  &=& a \, b \, \Big( \sum_{M(a) < i \leq m(b)} x_i^k 
        - \sum_{M(a) \leq i< m(b)} y_i^k \Big)   \qquad \mbox{if} \quad M(a) < m(b) \, , 
\ee
and $a \circ_k b = 0$ if $M(a) \geq m(b)$, where $a,b$ are any monomials in $\mathcal{A}$. 
This yields again a WNA algebra $\A$ (over $\mathbb{Z}$). 
 From $\A(f)$ one obtains a `supersymmetric' \cite{Stem85} 
(or `bisymmetric' \cite{MNR81}) version of quasi-symmetric functions. 
As a consequence of the new rules, (\ref{qsf-p_n}) is replaced by 
\be
    p_n \equiv f \circ_n f = \sum_i (x_i^n - y_i^n) \; .
\ee
Proposition~\ref{prop:qs_delta=p} also holds in this case. Together with 
$\delta_n(f) = p_n$, this turns (\ref{algebraicFay}) (or (\ref{functionalKP-id})) 
into `supersymmetric KP identities'. 
Such identitities in particular arise from a formal power series ansatz to 
solve the potential KP hierarchy 
\cite{Okhu+Wada83,Kupe00,Pani01,DMH04ncKP,DMH05KPalgebra}.

\section{A nonassociative hierarchy}
\label{section:nahier}
\setcounter{equation}{0}
Let $\A$ be a WNA algebra over the ring $\mathcal{C}^\infty$ of real 
or complex smooth functions of independent variables $\mathbf{t} = (t_1,t_2,\ldots)$. 
The equations
\be
  \pa_{t_n}(f) = f \circ_n f  \qquad \quad n=1,2, \ldots     \label{na_hier}
\ee
then constitute a `nonassociative hierarchy' according to the following proposition. 
We shall assume that $f \not\in \A'$, since otherwise (\ref{na_hier}) would 
reduce to a single equation. 
In the case of $\A_{\mathrm{free}}(f)$, the equations (\ref{na_hier}) 
are all independent. Depending on the choice of $\A$, there will be relations 
among them, in general. In the following, $\K$ stands for $\R$ or $\C$, and
$\A(f,\mathbb{K})$ denotes the WNA algebra generated in $\A$ by $f \in \A$ with 
coefficients in $\mathbb{K}$.

\begin{proposition}
\label{prop:na_hier}  \quad \\ 
(1) The flows (\ref{na_hier}) commute. \\
(2) For any solution $f$ of (\ref{na_hier}), $\delta_n(f) := f \circ_n f$ 
determines derivations $\delta_n$ of $\A(f,\mathbb{K})$. Hence
\be
     \pa_{t_n} = \delta_n  \quad \mbox{on} \; \A(f,\mathbb{K}) 
        \qquad \quad n=1,2, \ldots \; .  \label{na_hier-delta}
\ee
\end{proposition}
{\it Proof:} 
Since (\ref{na_hier}) implies $f_{t_n} := \pa_{t_n}(f) \in \A'$, it follows 
that the flow derivatives $\pa_{t_n}$ act as derivations of the products 
$\circ_m$ in $\A(f)$. The proof is analogous to that of 
proposition~\ref{prop:deriv-circ_n}. 
The commutativity of the flows can now be checked directly as follows, 
\bez
  (f_{t_m})_{t_n} &=& (f \circ_m f)_{t_n}
                   = f_{t_n} \circ_m f + f \circ_m f_{t_n}
                   = (f \circ_n f) \circ_m f + f \circ_m (f \circ_n f) \\
                  &=& f \circ_n (f \circ_m f) - f \circ_{m+n} f 
                      + (f \circ_m f) \circ_n f + f \circ_{m+n} f  \\
                  &=& (f \circ_m f) \circ_n f + f \circ_n (f \circ_m f) 
                   = (f_{t_n})_{t_m} \, , 
\eez
by use of proposition~\ref{prop:circ_m_circ_n-ass}. 

Since $\pa_{t_n}$ extends as a derivation to $\A(f,\mathbb{K})$ (where the coefficients 
of monomials in $f$ are independent of $\mathbf{t}$), (\ref{na_hier}) guarantees 
the consistency of extending $\delta_n(f) := f \circ_n f$ to $\A(f,\mathbb{K})$ 
by requiring the derivation property. 
\hfill $\square$
\vskip.2cm

The following proposition provides us with a formal solution of the 
initial value problem for (\ref{na_hier}), at least for a subclass of 
WNA algebras.

\begin{proposition}
\label{prop:S} 
Let $\A$ be a WNA algebra over $\K[[\mathbf{t}]]$, 
$f_0 \in \A \setminus \A'$ constant and generating a $\delta$-compatible subalgebra 
$\A(f_0,\K)$. Then
\be
    f := S(f_0) \qquad
    \mbox{with} \quad S := \exp\Big( \sum_{n \geq 1} t_n \, \delta_n \Big) 
           \label{S}
\ee
(where the $\delta_n$ are defined in terms of $f_0$) 
satisfies the nonassociative hierarchy (\ref{na_hier}). 
\end{proposition}
{\it Proof:} Since the $\delta_n$ are commuting derivations with respect to all 
the products $\circ_m$, $m=1,2, \ldots$, the linear operator $S$ on $\A(f_0,\K)$ is 
an automorphism with respect to all these products (which are defined via 
(\ref{circ_n}) in terms of $f_0$). Hence
\bez
 \pa_{t_n}(f) = \pa_{t_n} S(f_0)
             = S(\delta_n(f_0))
             = S(f_0 \circ_n f_0)
             = S(f_0) \circ_n S(f_0) 
             = f \circ_n f \; .
\eez
Since $\delta_n(f) \in \A(f_0)'$, we have $f - f_0 \in \A(f_0)'$, 
and thus $[f] = [f_0] \in \A(f_0)/\A(f_0)'$. 
Recalling proposition~\ref{prop:circ_n-equiv}, the product $\circ_n$ is 
equivalently defined in terms of $f$. This proves our assertion. 
Note that since $S(\delta_n(f_0)) = \delta_n (S(f_0)) = \delta_n(f)$, 
the derivation $\delta_n$ defined via (\ref{delta_n-def}) in terms of $f_0$ 
actually coincides with $\delta_n$ defined in terms of $f$. 
\hfill $\square$
\vskip.2cm

\noindent
{\bf Remark.} If $f \in \A \setminus \A'$ solves (\ref{na_hier}), then 
$\A(f,\mathbb{K})$ is $\delta$-compatible according to proposition~\ref{prop:na_hier}. 
If $\mathbf{t}'$ is another set of variables, and $S'$ the expression for $S$ 
with $\mathbf{t}$ replaced by $\mathbf{t}'$, then by use of (\ref{na_hier-delta}) 
we have $(S'(f))(\mathbf{t}) = f(\mathbf{t}+\mathbf{t}')$. 
\hfill $\square$
\vskip.2cm

For any $\nu \in \A \setminus \A'$ with $[\nu] = [f_0]$, the solution given 
by proposition~\ref{prop:S} has the property 
\be
    f = \nu - \phi  \qquad \mbox{with} \quad \phi \in \A' \; .    \label{f_decomp}
\ee
Inserting such a decomposition in (\ref{na_hier}), and assuming that $\nu$ is constant, 
turns the nonassociative hierarchy into
\be
    \phi_{t_n} = \nu \circ_n \phi + \phi \circ_n \nu - \phi \circ_n \phi 
                 - \nu \circ_n \nu
        \qquad \quad n=1,2, \ldots  \; .      \label{na_hier-phi}
\ee
Since the products $\circ_n$ are constructed in terms of the constant element $\nu$, 
these equations are of Riccati type. If the constant term in (\ref{na_hier-phi}) 
vanishes, they are of Bernoulli type (a simplifying restriction imposed in 
section~\ref{section:solutions}). Note that (\ref{na_hier-phi}) is a set of 
ordinary differential equations in the associative algebra $\A'$. According to 
proposition~\ref{prop:WNA}, we can formulate it in principle without any reference 
to a WNA algebra,
\be
   \phi_{t_1} &=& - g + L(\phi) + R(\phi) - \phi^2  \, ,   \label{nah_gLR1} \\
   \phi_{t_2} &=& - L(g) + R(g) + L^2(\phi) - R^2(\phi) + g \, \phi - \phi \, g 
                  - [ \phi \, L(\phi) - L(\phi) \, \phi ] \, ,   \label{nah_gLR2} \\
   \phi_{t_3} &=& - L^2(g) + LR(g) 
                  + L^3(\phi) - L(g) \, \phi 
                  + R(g) \, \phi - g \, L(\phi) 
                  + \phi \, [ L(g) - R(g) ]  \nonumber \\
              & & + R^3(\phi) - R(\phi) \, g 
                  - \phi \, L^2(\phi) + \phi \, g \, \phi 
                  + R(\phi) \, L(\phi) - R^2(\phi) \, \phi \, ,    \label{nah_gLR3}
\ee
and so forth. Here $\phi$ and $g$ are elements of an associative algebra $\mathcal{A}$ 
($=\A'$), $g$ constant, and $L,R$ commuting linear maps $\mathcal{A} \rightarrow \mathcal{A}$, 
which also commute with the partial derivatives $\pa_{t_n}$ and 
satisfy $L(ab)=L(a)b$, $R(ab)=aR(b)$ for all $a,b \in \mathcal{A}$. 
However, without reference to a WNA algebra structure, the building law 
underlying these equations is hard to detect. The combinatorics behind it is 
conveniently expressed in terms of a WNA algebra. 
\vskip.2cm

\begin{proposition}
Any solution $\phi$ of (\ref{nah_gLR1})-(\ref{nah_gLR3}) in an associative algebra 
$\mathcal{A}$ also solves the potential KP equation (with dependent variable in 
$\mathcal{A}$).
\end{proposition}
{\it Proof:} 
Use $g,L,R$ to define a WNA algebra $\A$ with $\A' = \mathcal{A}$. Let us 
denote the augmenting element by $\nu$. It follows that 
$f := \nu - \phi$ satisfies the first three equations of the nonassociative 
hierarchy (\ref{na_hier}). 
According to proposition~\ref{prop:na_hier}, there exist derivations $\delta_n$, 
$n=1,2,3$, of $\A(f,\mathbb{K})$, which have the properties stated in 
section~\ref{section:derivations}. But we know already from the introduction, 
that these derivations satisfy an identity which via (\ref{na_hier}) (for $n=1,2,3$) 
shows that $-f$ solves the potential KP equation. Since $\nu$ is constant, and 
since the potential KP equation does not contain the dependent variable without 
derivatives acting on it, it follows that $\phi$ solves it. 
\hfill $\square$
\vskip.2cm

In principle we can extend (\ref{nah_gLR1})-(\ref{nah_gLR3}) by translating further 
equations of (\ref{na_hier-phi}). Solutions of this extended system then yield 
solutions of the respective part of the potential KP hierarchy. 
But only when expressed in terms of a WNA structure the whole set of 
equations for $\phi$ takes a simple and compact form. 
In the following subsection it will indeed be proved  
that the whole potential KP hierarchy is a consequence of (\ref{na_hier-phi}). 
The splitting off of a constant term in (\ref{f_decomp}) is then quite natural 
from the point of view that the potential $\phi$ is obtained from the proper 
KP variable by integration with respect to $t_1$. Thus $\nu$ plays the 
role of a constant of integration. 
In another subsection we show that also the Burgers hierarchy results 
from the nonassociative hierarchy (\ref{na_hier}), by restriction to a 
certain subclass of WNA algebras.

\subsection{Relation with the KP hierarchy}
\label{subsection:KP}
We will make use of the following notation, which can be traced back 
at least to \cite{Sato+Sato82}. For any object $F$ depending smoothly 
on $\mathbf{t}=(t_1,t_2,\ldots)$ let 
\be
      F_{[\la]}(\mathbf{t}) := F(\mathbf{t}+[\la]) \, , \qquad
      F_{-[\la]}(\mathbf{t}) := F(\mathbf{t}-[\la])  \, , 
\ee
where $[\la] := (\la, \la^2/2, \la^3/3, \ldots)$.

\begin{proposition}
Let $\A$ be a WNA algebra over $\mathcal{C}^\infty$, 
$\nu \in \A \setminus \A'$ constant, and $\phi \in \A'$ 
a solution of the \emph{ordinary} differential equations (\ref{na_hier-phi}). 
Then $\phi$ also solves the potential KP hierarchy, for which a 
functional representation is given by \cite{Bogd+Kono98,Bogd99}
\be
   \sum_{i,j,k=1}^3 \varepsilon_{ijk} \, \left( \la_i^{-1} + \phi_{[\la_k]} - \phi \right)
           \left( \phi_{[\la_i]} - \phi \right)_{[\la_k]} = 0  
          \; .  \label{functionalKP}
\ee
\end{proposition}
{\it Proof:} 
Setting $f := \nu - \phi$ satisfies the nonassociative hierarchy (\ref{na_hier}). 
By use of (\ref{na_hier-delta}), in $\A(f,\mathbb{K})$ we have
\bez
    H_n(f) = \chi_n(\tilde{\pa})(f) \, , \qquad
    E_n(f) = (-1)^n \, \chi_n(-\tilde{\pa})(f) \, ,    \label{H_n=chi(pa)}
\eez
and thus
\bez
    H(\la)(f) = f_{[\la]} \, , \qquad 
    E(\la)(f)) = f_{-[-\la]} \; . 
\eez
This turns the identity (\ref{functionalKP-id}) into (\ref{functionalKP}). 
\hfill $\square$
\vskip.2cm

We have shown that the potential KP hierarchy is a consequence of the 
`nonassociative hierarchy' (\ref{na_hier}), assuming the decomposition (\ref{f_decomp}). 
In particular, the $\phi$ determined via (\ref{f_decomp}) by proposition~\ref{prop:S} 
provides us with a (formal) solution of the potential KP hierarchy (see also 
section~\ref{subsection:free-to-KP}). 
\vskip.2cm

\noindent
{\bf Remark.} In the limit $\la_3 \to 0$, (\ref{functionalKP}) leads to
\be
     \Big( \phi_{[\la_2]} - \phi_{[\la_1]} \Big)_x
 &=& \Big( \la_1^{-1} - \la_2^{-1} + \phi_{[\la_2]} - \phi_{[\la_1]} \Big) \,
    (\phi_{[\la_1]+[\la_2]} - \phi_{[\la_1]} - \phi_{[\la_2]} + \phi) \nonumber \\
  & & - [\phi_{[\la_1]}-\phi , \phi_{[\la_2]}-\phi]
      \label{pre-Fay}
\ee
where $x := t_1$. Alternatively, this is obtained from (\ref{algebraicFay}) by 
use of (\ref{na_hier}) and (\ref{f_decomp}). 
$\phi$'s without derivatives acting on them drop out of this formula (as obvious 
from its origin). 
Representing $\A'$ by a \emph{commutative} algebra of functions, such that 
$\phi \mapsto \tau_x/\tau$ with a function $\tau$, (\ref{pre-Fay}) becomes
\be
   \Big( \log \Big( \la_1^{-1} - \la_2^{-1}
   + \frac{\tau_{[\la_2],x}}{\tau_{[\la_2]}}
   - \frac{\tau_{[\la_1],x}}{\tau_{[\la_1]}} \Big) \Big)_x
  = \left( \log{\tau_{[\la_1]+[\la_2]} \tau \over \tau_{[\la_1]} \tau_{[\la_2]}}
    \right)_x \, ,     \label{Fay-inter}
\ee
and integration yields
\be
    ( \la_1^{-1} - \la_2^{-1} ) \, \tau_{[\la_1]} \tau_{[\la_2]}
    + \tau_{[\la_1]} \tau_{[\la_2],x} - \tau_{[\la_2]} \tau_{[\la_1],x}
    = C \, \tau_{[\la_1]+[\la_2]} \, \tau \, ,
\ee
with a constant of integration $C$. 
Fixing the latter such that the terms with negative powers of $\la_1$ or $\la_2$ 
drop out in the formal series, the result is
\be
    \tau_{[\la_1]} \, \tau_{[\la_2],x} - \tau_{[\la_1],x} \, \tau_{[\la_2]} 
  = ( \la_1^{-1} - \la_2^{-1} ) (\tau_{[\la_1] + [\la_2]} \, \tau
    - \tau_{[\la_1]} \, \tau_{[\la_2]}) \, , 
    \label{diffFay}
\ee
which is known as the \emph{differential Fay identity} 
 \cite{Shio86,Adle+vanM94,Adle+vanM99,Taka+Take95}. 
It is equivalent to the whole KP hierarchy (with dependent variable in $\mathcal{C}^\infty$) \cite{Taka+Take95}. 
(\ref{pre-Fay}) should thus be regarded as a `noncommutative' version of the 
differential Fay identity.
Expansion of (\ref{pre-Fay}) indeed yields to lowest nonvanishing order 
the `noncommutative' potential KP equation
\be
   ( \phi_t - \frac{1}{4} \phi_{xxx} - \frac{3}{2} \phi_x{}^2 )_x
   = \frac{3}{4} \phi_{yy} + \frac{3}{2} [\phi_y , \phi_x ] \, , 
\ee
where $y := t_2$ and $t := t_3$ (cf. \cite{DMH04ncKP}, for example).
\hfill $\square$

\subsection{Relation with Burgers hierarchies}
\label{subsection:Burgers}
If $f$ solves the nonassociative hierarchy (\ref{na_hier}), so that 
(\ref{na_hier-delta}) holds, then the identities (\ref{HE-recur}) in 
$\A(f,\mathbb{K})$ lead to
\be
   \chi_{n+1}(\tilde{\pa}) f = f \, \chi_n(\tilde{\pa}) f
   \qquad n=0,1,2, \ldots \, ,    \label{pre-left-Burgers}
\ee
and also 
\be
   \chi_{n+1}(-\tilde{\pa}) f = - ( \chi_n(-\tilde{\pa}) f ) \, f
   \qquad n=0,1,2, \ldots  \; .   \label{pre-right-Burgers}
\ee
Assuming the decomposition (\ref{f_decomp}) with constant $\nu$, 
the $n=0$ equation reads
\be
   \phi_x = - \nu^2 + \nu \, \phi + \phi \, \nu - \phi^2 \, , 
\ee
and the remaining equations form the hierarchies
\be
   \chi_{n+1}(\tilde{\pa}) \phi = (\nu - \phi) \, \chi_n(\tilde{\pa}) \phi
   \qquad n=1,2, \ldots \, ,    \label{na-left-Burgers}
\ee
respectively
\be
   \chi_{n+1}(-\tilde{\pa}) \phi = (\chi_n(-\tilde{\pa}) \phi) \, (\phi - \nu)
   \qquad n=1,2, \ldots  \; .   \label{na-right-Burgers}
\ee
Choosing the WNA algebra $\A$ such that 
$\nu \, a =0$, respectively $a \, \nu =0$, for all $a \in \A'$,
the two hierarchies take the form
\be
      \chi_{n+1}(\tilde{\pa}) \phi =  - \phi \, \chi_n(\tilde{\pa}) \phi
   \qquad n=1,2, \ldots \, ,    \label{left-Burgers}
\ee
respectively
\be
   \chi_{n+1}(-\tilde{\pa}) \phi = (\chi_n(-\tilde{\pa}) \phi) \, \phi
   \qquad n=1,2, \ldots  \; .   \label{right-Burgers}
\ee
These are the two versions of the Burgers hierarchy with dependent 
variable $\phi$ in a noncommutative associative algebra. 
For $n=1$, we have
\be
    \phi_y + \phi_{xx} = - 2 \, \phi \, \phi_x \, ,
\ee
respectively
\be
    \phi_y - \phi_{xx} = 2 \, \phi_x \, \phi \, ,
\ee
which are indeed the `left' and `right' versions of the 
`noncommutative' \emph{Burgers equation} (see \cite{LRB83,Svin89,Hama+Toda03b,Kupe05}, 
for example). 
The sets of equations (\ref{left-Burgers}) and (\ref{right-Burgers}) 
can be expressed in the compact form
\be
  (\la^{-1} + \phi)_x = (\la^{-1} + \phi)( \phi_{[\la]} - \phi ) \, ,
\ee
respectively
\be
  (\la^{-1} - \phi)_x = ( \phi_{-[\la]} - \phi )(\la^{-1} - \phi) \; .
\ee
Representing $\A'$ by a \emph{commutative} algebra of functions, 
such that $\phi \mapsto \tau_x/\tau$ with a function $\tau$, these equations can be 
integrated. Choosing the integration constant in such a way that the 
terms with negative powers of $\la$ drop out of the formal series, we obtain 
the linear functional equations $\tau_x = \la^{-1} (\tau_{[\la]} - \tau)$, 
respectively $\tau_x = \la^{-1} (\tau - \tau_{-[\la]})$.

\section{From solutions of the nonassociative hierarchy to solutions of KP hierarchies}
\label{section:solutions}
\setcounter{equation}{0}
At least for a restricted class of WNA algebras we can achieve a much simpler 
form of the whole nonassociative hierarchy. 
The trick is to introduce an `auxiliary product' in terms of which we 
can `resolve' the products $\circ_n$ in $\A'$ to a simple form. 
This is done in subsection~\ref{subsection:simpl}. 
In the case where $\A'$ is a matrix algebra, it leads to solutions of 
matrix potential KP hierarchies in subsection~\ref{subsection:KPsol}. 
These in turn lead to solutions of the scalar potential KP hierarchy. 
In subsection~\ref{subsection:free-to-KP} another route is taken. Starting 
from a free associative algebra, we construct a WNA algebra to which 
proposition~\ref{prop:S} can be applied. This leads to a kind of 
`universal solution' of the nonassociative hierarchy from which the 
matrix KP hierarchy solutions of subsection~\ref{subsection:KPsol} 
are recovered by homomorphisms from $\A'$ to the corresponding matrix algebra.

\subsection{Simplifying the nonassociative hierarchy}
\label{subsection:simpl}
Let $(\mathcal{A},\ci)$ be any associative algebra over $\mathcal{C}^\infty$ 
(which is the ring of real or complex functions of $t_1,t_2,\ldots$), 
and $L,R$ commuting linear maps such that $L(a \ci b)=L(a) \ci b$ and 
$R(a \ci b)=a \ci R(b)$ for all $a,b \in \mathcal{A}$. 
The notation $La := L(a)$ and $aR := R(a)$, used in the following, 
shall remind us of these properties. 
A new associative product in $\mathcal{A}$ is then given by
\be
    a \circ_1 b := (a R) \ci b - a \ci (L b) \; .  \label{new_product}
\ee
Augmenting $(\mathcal{A},\circ_1)$ with an element $\nu$ such that
\be
    \nu \circ_1 \nu := 0 \, , \qquad
    \nu \circ_1 a := L a \, , \qquad
    a \circ_1 \nu := - a R \, ,     \label{nu-circ_1}
\ee
we obtain a WNA algebra $(\A,\circ_1)$ with the property 
$\A' = \mathcal{A}$. Restricted to $\A'$, we have $L_\nu = L$ and $R_\nu = -R$. 
For the products (\ref{circ_n}), defined with $\nu$, 
one easily proves by induction that
\be
  &&  \nu \circ_n \nu = 0 \, , \qquad
      \nu \circ_n a = L^n a \, ,
             \qquad
      a \circ_n \nu = - a R^n \, , \nonumber \\
  &&  a \circ_n b
      = \sum_{k=0}^{n-1} (-1)^k \, (R_{\nu}^k a) \circ_1 L_{\nu}^{n-k-1} b
      = (a R^n) \ci b - a \ci L^n b  \label{circ_n-LR}
\ee
for all $a,b \in \mathcal{A}$. Note that in the last equation 
we have a telescoping sum as a consequence of (\ref{new_product}). 
The complicated formula for $a \circ_n b$ in $(\A,\circ_1)$ 
is thus drastically simplified when expressed in terms of the 
product $\ci$. 
As a consequence, (\ref{na_hier-phi}) can be written as
\be
    \phi_{t_n} = L^n \phi - \phi R^n + \phi \ci L^n \phi - \phi R^n \ci \phi 
     \qquad \quad  n=1,2,\ldots  \; .    \label{nahier-phi-ci}
\ee
According to our general results, \emph{any} solution of this system is a 
solution of the potential KP hierarchy in $(\mathcal{A},\circ_1)$. 
\vskip.2cm

\noindent
{\bf Remark.} If we generalize the first equation of (\ref{nu-circ_1}) to 
$\nu \circ_1 \nu := g$ with some $g \in \mathcal{A}$, 
the resulting expressions for $\circ_n$-products are much more complicated. 
In particular, we obtain 
\be
  &&  \nu \circ_2 \nu = L g + g R \, , \quad
      \nu \circ_2 a = L^2 a - (g R) \ci a + g \ci (L a) \, ,  \nonumber \\
  &&  a \circ_2 \nu = (a R) \ci g - a \ci (L g) - a R^2 \, , \quad
      a \circ_2 b = (a R^2) \ci b - a \ci (L^2 b) \, ,
\ee
and 
\be
   && \nu \circ_3 \nu = L^2 g + L g R + g R^2 + g \ci (Lg) - (gR) \ci g \, ,  \nonumber \\
   && \nu \circ_3 a = L^3 a + g \ci (L^2 a) + (Lg) \ci (L a) - (LgR) \ci a - (g R^2) \ci a  
       \, ,   \nonumber \\
   && a \circ_3 \nu = - a R^3 + (a R^2) \ci g + (aR) \ci (gR) - a \ci (LgR) - a \ci (L^2 g)
         \, ,  \nonumber \\
   && a \circ_3 b = (a R^3) \ci b - a \ci (L^3 b) 
          - (aR) \ci ( (gR) \ci b - g \ci (Lb) ) 
          + a \ci L( (gR) \ci b - g \ci (Lb) )  \, , \qquad \quad   \label{circ_3-gLR}
\ee
for all $a,b \in \mathcal{A}$. 
The first two equations of (\ref{na_hier-phi}) then read 
\be
    \phi_{t_1} &=& -g + L \phi - \phi R - (\phi R) \ci \phi + \phi \ci (L \phi) \, , \\
    \phi_{t_2} &=& - L g - g R + L^2 \phi - (g R) \ci \phi + g \ci (L \phi) \nonumber \\
    && + (\phi R) \ci g - \phi \ci (L g) - \phi R^2 - (\phi R^2) \ci \phi 
       + \phi \ci (L^2 \phi) \, ,
\ee
and by use of (\ref{circ_3-gLR}) the third equation already results in a rather lengthy 
expression. If we can solve these ordinary differential equations, then a solution $\phi$ 
also solves the potential KP equation in $(\mathcal{A},\circ_1)$. 
For $g \neq 0$, these Riccati-type equations are difficult to solve, however. 
\hfill $\square$

\subsection{Solutions of matrix potential KP hierarchies}
\label{subsection:KPsol}
In the framework of section~\ref{subsection:simpl}, let 
$\mathcal{A} = \mathcal{M}(M,N)$, the algebra of $M \times N$ matrices with the product
\be
    A \ci A' := A K A'    \qquad \quad  \forall A,A' \in  \mathcal{M}(M,N) \, , 
\ee
where $K$ is a fixed $N \times M$ matrix (and concatenation means the 
ordinary matrix product). Furthermore, for the linear maps $L,R$ we 
choose the multiplication operators by an $M \times M$ matrix $L$, 
and an $N \times N$ matrix $R$ (using the same symbols, for simplicity). 
Now the Bernoulli-type equations (\ref{nahier-phi-ci}) can be easily solved. 
In terms of $\tilde{\phi} := e^{-\xi(L)} \, \phi \, e^{\xi(R)}$, where
\be
  \xi(L) := \sum_{n \geq 1} t_n \, L^n \, , \qquad \xi(R) := \sum_{n \geq 1} t_n \, R^n \, ,
\ee
it takes the form 
$\tilde{\phi}_{t_n} = \tilde{\phi} \, (e^{-\xi(R)} \, K \, e^{\xi(L)})_{t_n} \tilde{\phi}$, 
which is solved by 
$\tilde{\phi} = (I_M -B)^{-1} C$ (provided the inverse exists) with a constant 
matrix $C \in \mathcal{M}(M,N)$, the $M \times M$ unit matrix $I_M$, and 
\be
      B := C \, e^{-\xi(R)} \, K \, e^{\xi(L)} \; .
\ee
This leads to 
\be
    \phi = e^{\xi(L)} \, (I_M - B)^{-1} \, C \, e^{-\xi(R)} \, , 
\ee
which can also be expressed as
\be
   \phi = \phi^{(1)} \, (I_N - K \, \phi^{(1)})^{-1} 
        = (I_M - \phi^{(1)} \, K)^{-1} \, \phi^{(1)}  \, , \quad
   \phi^{(1)} := e^{\xi(L)} \, C \, e^{-\xi(R)} \; .
           \label{phi-solution-alt}
\ee

 From the general theory, we know that $\phi$ solves the potential KP hierarchy 
in $(\mathcal{M}(M,N),\circ_1)$, where the product is now given by 
\be
     A \circ_1 A' = A \, (R K - K L) \, A'  \; .
\ee

In the following, we show how to obtain solutions of the scalar potential KP hierarchy 
from the above solutions of a matrix potential KP hierarchy (which is in the spirit 
of the `operator approach' to solutions of soliton equations \cite{March88,Carl+Schi99}). 
Let us choose the matrices $K,L,R$ such that
\be
    R K - K L = v u^T \, ,     \label{RKLv}
\ee
with nonvanishing $v \in \K^N$, $u \in \K^M$, and $u^T$ denotes 
the transpose of $u$. Then we have 
\be
      A \circ_1 A' = A \, v u^T A' \, ,
\ee 
and the linear map $\Psi : \mathcal{M}(M,N) \rightarrow \mathcal{C}^\infty$, defined by
\be
    \Psi(A) := u^T \, A \, v  \, , 
\ee
is an algebra homomorphism, i.e.
\be
    \Psi(A \circ_1 A') = u^T \, A \, v \, u^T \, A' \, v = \Psi(A) \, \Psi(A') \; .
\ee
As a consequence, if $\phi$ solves the potential KP hierarchy in $(\mathcal{M}(M,N),\circ_1)$, 
then $\Psi(\phi)$ solves the scalar potential KP hierarchy. 
\vskip.2cm

\noindent
{\bf Remark.} 
Equation (\ref{RKLv}) amounts to $\mathrm{rank}(R K - K L) = 1$, a condition 
which also appeared in \cite{Carl+Schi99} within an analysis of the KP equation, 
more generally for operators on a Banach space (see also \cite{Gant59,GHS06}). 
Let $v$ be an eigenvector of $R$ to an eigenvalue $\kappa$, and $K = v \, u^T$ with 
any vector $u \neq 0$. Then $R K - K L = v \, ( \kappa \, u^T - (L^T u)^T )$ is 
of rank $1$ unless the right hand side vanishes. To avoid this, it is sufficient to 
have any $u \neq 0$ which is \emph{not} an eigenvector of $L^T$ to the eigenvalue $\kappa$. 
Thus we have to exclude the case $L = \kappa \, I_M$. It follows that 
$\mathrm{rank}(R K - K L) = 1$ always has a solution except when $R= \kappa \, I_N$ 
and $L = \kappa \, I_M$ with $\kappa \in \K$. Note, however, that the most 
interesting solutions are those for which $K$ has maximal rank. 
\hfill $\square$ 
\vskip.2cm

For the solution $\phi$ obtained above, we find
\be
   \Psi(\phi) &=& u^T \, e^{\xi(L)} \, (I_M - B)^{-1} \, C \, e^{-\xi(R)} \, v 
               = \tr( e^{\xi(L)} \, (I_M - B)^{-1} \, C \, e^{-\xi(R)} \, v \, u^T )  \nonumber \\
              &=& \tr( (I_M - B)^{-1} \, C \, e^{-\xi(R)} \, (RK-KL) \, e^{\xi(L)} )  
               = - \tr( (I_M-B)^{-1} \, B_x )  \nonumber \\
              &=& \tr(\log(I_M - B))_x
               = ( \log \tau )_x  \, , 
\ee
with $x := t_1$ and
\be
     \tau := \det(I_M - B)  \; .
\ee
\vskip.2cm

\noindent
{\bf Example 1.} 
Setting  $M=N$, $L = \mathrm{diag}(\mathrm{p}_1,\ldots,\mathrm{p}_N)$,
$R = \mathrm{diag}(\mathrm{q}_1,\ldots,\mathrm{q}_N)$ with entries in $\K$,
$u=v$ with $v^T = (1,1,\ldots,1)$, the relation (\ref{RKLv}) leads to
\be
    K_{ij} = (\mathrm{q}_i - \mathrm{p}_j)^{-1} \, ,
\ee
assuming $\mathrm{q}_i \neq \mathrm{p}_j$ for $i,j=1,\ldots,N$. 
Choosing $C = \mathrm{diag}(c_1, \ldots, c_N)$, $\Psi(\phi)$ becomes the $N$-soliton 
solution of the potential KP hierarchy. 
For example, with $N=2$ and $c_1 = (p_1-q_1) e^{\alpha_1}, c_2 = (p_2-q_2) e^{\alpha_2}$,  
where $\alpha_1,\alpha_2 \in \K$, we obtain the tau function of a 2-soliton 
solution in the form (cf. \cite{Hiro04}, for example)
\be
    \tau = 1 + e^{\eta_1} + e^{\eta_2}
           + \frac{(q_2-q_1)(p_2-p_1)}{(q_2-p_1)(p_2-q_1)} \, e^{\eta_1+\eta_2} \, , 
\ee
where $\eta_i := \alpha_i + \xi(p_i) - \xi(q_i)$, $i=1,2$, and 
$\xi(q) = \sum_{n \geq 1} t_n \, q^n$. 
\hfill$\square$
\vskip.2cm

\noindent
{\bf Example 2.} For nilpotent matrices $L,R$, the function $\tau$ is \emph{polynomial} 
in the independent variables. Let us set
\be
    L = \left(\begin{array}{cccc} 0 & 1 & 0 & 0 \\
    0 & 0 & 1 & 0 \\ 0 & 0 & 0 & 1 \\ 0 & 0 & 0 & 0 \end{array}\right) \, ,
    \qquad
    R = \left(\begin{array}{cc} 0&0\\1&0\end{array}\right) \; .
\ee
Then (\ref{RKLv}) is satisfied by setting $v^T=(1,1)$, $u^T = (0,0,1,1)$, and
\be
    K = \left(\begin{array}{cccc} 0 & -1 & -1 & 1 \\
    -1 & -2 & 0 & 0 \end{array}\right) \; .
\ee
Choosing furthermore 
\be
    C = \left(\begin{array}{cc} 1+2\alpha^{-1}\beta & 2\alpha  \\
    0 & -\alpha \\ \beta & \alpha  \\
    \beta & 0 \end{array}\right)
\ee
with $\alpha, \beta \in \K$, we obtain 
\be
    \tau = \det(I - B)
  = 1 + \alpha \left(t_2 - {1 \over 2} t_1^2 \right)
    + \beta \left(t_2 + {1 \over 2} t_1^2 \right)
    + \alpha \beta \left( - t_1 t_3 + t_2^2 + {1 \over 12} t_1^4 \right) \, ,
\ee
which is indeed a special tau function of the KP hierarchy (see, for example, 
\cite{MJD00}, p. 59).
\hfill$\square$
\vskip.2cm

\noindent
{\bf Remark.} 
Since we obtained common solutions of all equations of the
nonassociative hierarchy (\ref{na_hier}), setting the matrix $L=0$, respectively 
the matrix $R =0$, they also solve the Burgers hierarchy (\ref{left-Burgers}), 
respectively (\ref{right-Burgers}). In particular, the KP multi-soliton 
solutions become with these restrictions also solutions of the Burgers hierarchies.
\hfill$\square$
\vskip.2cm

\noindent
{\bf Remark.} 
Let $V$ be an $N \times P$ and $U$ an $M \times P$ matrix such that
$R K - K L = V \, U^T$ (which generalizes (\ref{RKLv})). 
Then we can define $\Psi$ more generally as a map 
$\mathcal{M}(M,N) \rightarrow \mathcal{M}(P,P)$, with the ordinary 
matrix product in $\mathcal{M}(P,P)$, 
by setting $\Psi(A) := U^T A V$ for all $A \in \mathcal{M}(M,N)$. 
This yields indeed an algebra homomorphism since 
$\Psi(A \circ_1 A') = (U^T A \, V)(U^T A' \, V)$.
As a consequence, $\Psi(\phi)$ with the above solution $\phi$ 
solves the $P \times P$ matrix potential KP hierarchy (with the ordinary 
matrix product). 
\hfill$\square$

\subsection{From a free algebra to solutions of KP hierarchies}
\label{subsection:free-to-KP}
In this subsection we demonstrate in particular how the solutions of the matrix 
potential KP hierarchies obtained in the previous subsection can be obtained 
via application of proposition~\ref{prop:S}. 
We choose $\mathcal{A} = \mathcal{A}_{\mathrm{free}}$, the \emph{free associative} 
algebra considered in section~\ref{subsection:freeWNA}, now taken over $\mathcal{C}^\infty$. 
For the maps $L,R$ we have (using the notation of section~\ref{subsection:simpl}) 
\be
    L c_{r,s} = c_{r+1,s} \, , \qquad
    c_{r,s} R = c_{r,s+1} \, , \qquad
    c_{r,s} = L^r c \, R^s \; . 
\ee
Now $\ci$ denotes the product in $\mathcal{A}_{\mathrm{free}}$ (for which we 
previously used concatenation). 
Since $(\mathcal{A}_{\mathrm{free}},\ci)$ is free, it is consistent to define derivations 
$\delta_n$, $n=1,2, \ldots$, by 
\be
    \delta_n(c_{r,s}) := L^n c_{r,s} - c_{r,s} R^n = c_{r+n,s} - c_{r,s+n} 
          \label{delta_n(p)-LR}
\ee
and the derivation rule. 
They satisfy $\delta_n(c_{r,s}) = L^r \, \delta_n(c) \, R^s$ and obviously commute. 
Setting 
\be
        \delta_n(\nu) := 0 \, , 
\ee
the $\delta_n$ are also derivations of $(\A,\circ_1)$ (which is 
constructed according to section~\ref{subsection:simpl}), 
and then with respect to all the products $\circ_m$, $m=1,2,\ldots$ 
(cf. proposition~\ref{prop:deriv-circ_n}).

\begin{proposition}
\be
       \delta_n(c^{\ci m})
  &=& \nu \circ_n c^{\ci m} + c^{\ci m} \circ_n \nu
      - \sum_{k=1}^{m-1} c^{\ci k} \circ_n c^{\ci (m-k)}
                            \label{delta_n(p^circm)}
\ee
for $m,n = 1,2,\ldots$, where $c^{\ci \, n}$ denotes the $n$th
power of $c$ using the product $\ci$.
\end{proposition}
{\it Proof:} Applying the derivation rule, we find 
\bez
      \delta_n(c^{\ci m})
  &=& \sum_{k=0}^{m-1} c^{\ci k} \ci \delta_n(c) \ci c^{\ci (m-k-1)}
   =  \sum_{k=0}^{m-1} c^{\ci k} \ci ( L^n c - c R^n )
      \circ c^{\ci (m-k-1)}    \\
  &=&  \sum_{k=0}^{m-1} \Big( c^{\ci k} \ci L^n c^{\ci (m-k)}
        - (c^{\ci (k+1)} R^n) \ci c^{\ci (m-k-1)} \Big)  \\
  &=&  L^n c^{\ci m} - c^{\ci m} R^n 
      + \sum_{k=1}^{m-1} \Big( c^{\ci k} \ci L^n c^{\ci (m-k)}
        - (c^{\ci k} R^n) \ci c^{\ci (m-k)} \Big) \, , 
\eez 
which by use of (\ref{circ_n-LR}) is equal to the right hand
side of (\ref{delta_n(p^circm)}). 
\hfill $\square$ 
\vskip.2cm

As a consequence of (\ref{delta_n(p^circm)}), the formal power series 
(in a parameter $\epsilon$) 
\be
    f_0 := \nu - \phi_0 \, , \qquad
    \phi_0 := \sum_{n \geq 1} \epsilon^n \, c^{\ci \, n}   \label{f=nu-sump}
\ee
satisfies
\be
    \delta_n(f_0) = f_0 \circ_n f_0   \qquad n=1,2,\ldots 
                   \label{delta_n-id}
\ee
(which is (\ref{delta_n-def}) with $f$ replaced by $f_0$). 
Here we made use of the fact that, according to proposition~\ref{prop:circ_n-equiv}, 
the products $\circ_n$ only depend on the equivalence class $[f_0] = [\nu]$ in $\A/\A'$. 
\vskip.2cm

In conclusion, the subalgebra $\A(f_0)$ of $\A$ generated by $f_0$ is  
$\delta$-compatible. According to proposition~\ref{prop:S}, 
$f := S(f_0)$ then satisfies (\ref{na_hier}) and $\phi := \nu - f$ solves the 
KP hierarchy (cf. section~\ref{subsection:KP}). 
\vskip.2cm

Since, in the case under consideration, the operator $S$ defined in (\ref{S}) 
is also a homomorphism of the product $\ci$, we have 
\be
    \phi = S(\phi_0) = \sum_{n \geq 1} \epsilon^n \, S(c)^{\ci n} 
    \, ,        \label{phi=S(phi_0)}
\ee
where, by use of the definition of the derivations $\delta_n$, 
\be
     S(c) = e^{\xi(L)} \, c \, e^{-\xi(R)}  \; .
\ee
We thus arrived at a solution of the potential KP hierarchy in
$(\mathcal{A}_{\mathrm{free}},\circ_1)$.
Since the product $\circ_1$ in $\mathcal{A}_{\mathrm{free}}$ is given in terms
of the product $\ci$ of the free associative algebra,
solutions in other associative algebras are obtained as follows.
Let $\rho$ be a homomorphism from $(\mathcal{A}_{\mathrm{free}},\ci)$ to an
associative algebra $(\mathcal{A},\Cdot)$, and 
$\tilde{L}, \tilde{R} : \mathcal{A} \to \mathcal{A}$
commuting mappings such that $\tilde{L}(x \Cdot y) = (\tilde{L}x) \Cdot y$
and $(x \Cdot y) \tilde{R} = x \Cdot (y \tilde{R})$ for all $x,y \in \mathcal{A}$. 
Then $\mathcal{A}$ can be supplied with the new associative product
$x \diamond y := (x \tilde{R}) \Cdot y - x \Cdot (\tilde{L} y)$ (cf. (\ref{new_product})). 
Moreover, if $\rho$ has the properties $\rho(L a) = \tilde{L} \rho(a)$ and 
$\rho(a R) = \rho(a) \tilde{R}$, then $\rho$ satisfies 
\be
   \rho( a \circ_1 b ) = \rho(a) \diamond \rho(b) \, .
\ee
The `universal solution' given above determines a solution of
the potential KP hierarchy in $(\mathcal{A},\diamond)$. A class of examples
is presented in the following, which makes contact with results obtained
in the previous subsection.
\vskip.2cm

Let $\mathcal{M}(M,N)$ be again the algebra of $M \times N$ matrices 
with the product
\be
    A \Cdot A' := A K A' \, ,    \label{K-product}
\ee
where $K$ is a fixed $N \times M$ matrix. 
Furthermore, we choose an $M \times M$ matrix $L$, 
an $N \times N$ matrix $R$, and a matrix $C \in \mathcal{M}(M,N)$. In terms of these matrices 
we define an algebra homomorphism $\rho \, : \, (\mathcal{A}_{\mathrm{free}},\ci) \rightarrow (\mathcal{M}(M,N),\Cdot)$ by
\be
    \rho(c_{m,n}) := L^m C R^n \qquad \quad   m,n = 0,1,2, \ldots  
\ee
and 
\be
     \rho(a \ci b) = \rho(a) \Cdot \rho(b) 
     \qquad \quad \forall a,b \in \mathcal{A}_{\mathrm{free}}  \; . 
\ee
It follows that
\be  
    \rho(L a) = L \rho(a) \, , \qquad   \rho(a R) = \rho(a) R  
\ee
(where, on the left hand sides, $L$ and $R$ denote the linear maps introduced 
in the beginning of this subsection) and thus
\be
    \Phi^{(1)} := \rho( S(c) ) = e^{\xi(L)} \, C \, e^{-\xi(R)} \, , 
\ee
with the matrices $L,R$ on the right hand side. 
Now we obtain (see also the `trace method' \cite{Okhu+Wada83,Kupe00,Pani01,DMH04ncKP}) 
\be
   \Phi &:=& \rho(\phi) 
         = \sum_{n \geq 1} \epsilon^n \, (\Phi^{(1)})^{\Cdot n}
         = \sum_{n \geq 1} \epsilon^n \, \Phi^{(1)} \, (K \, \Phi^{(1)})^{n-1}  \nonumber \\
        &=& \epsilon \, \Phi^{(1)} \, (I_N - \epsilon \, K \, \Phi^{(1)})^{-1} 
         = \epsilon \, (I_M - \epsilon \, \Phi^{(1)} \, K)^{-1} \, \Phi^{(1)} \; . 
           \label{Phi-solution}
\ee
This coincides (for $\epsilon=1$) with the matrix solution obtained in 
section~\ref{subsection:KPsol} (see (\ref{phi-solution-alt})). 
Note that $\rho( a \circ_1 b ) = \rho(a) \, (RK-KL) \, \rho(b) =: \rho(a) \diamond \rho(b)$.
\vskip.2cm

\noindent
{\bf Remark.} 
If $U \in \mathcal{M}(M,P)$ and $V \in \mathcal{M}(N,P)$ are such that 
$R K - K L = V U^T$, then $\sigma : \A_{\mathrm{free}}' \rightarrow \mathcal{M}(P,P)$,  
defined by $\sigma(a) := U^T \rho(a) \, V$, is an algebra homomorphism, 
$\sigma(a \ci_1 b) = \sigma(a) \, \sigma(b)$, which sends $\phi$ to a solution 
$\sigma(\phi)$ of the $P \times P$ matrix potential KP hierarchy. 
But since $\sigma$ is only defined on the associative subalgebra $\A_{\mathrm{free}}'$, 
the information how to construct solutions, which is encoded in the WNA structure 
of $\A_{\mathrm{free}}$, is not carried over by $\sigma$ to the target algebra 
$\mathcal{M}(P,P)$. 
\hfill$\square$

\noindent
{\bf Remark.} 
For fixed $k \in \mathbb{N}$, the two-sided ideal $\mathcal{I}_k$ in 
$(\mathcal{A}_{\mathrm{free}},\ci)$, 
generated by $\{ \delta_k(c_{m,n}) = c_{m+k,n} - c_{m,n+k} \, | \, m,n=0,1,\ldots \}$ 
is invariant under the derivations $\delta_j$, $j=1,2, \ldots$. 
Then $\A_k := \A/\mathcal{I}_k$ is compatible with the derivations $\delta_n$. 
In the quotient algebra, we have $\delta_{nk} \equiv 0$ for all $n \in \mathbb{N}$, 
so that (\ref{na_hier}) requires $\phi_{t_{nk}}=0$. 
In this way contact is made with the Gelfand-Dickey reductions of the KP hierarchy 
\cite{Dick03}. The map $\rho$ projects to $\A_k$ if $L^k \, C = C \, R^k$. 
As a consequence, 
\be
    \exp\Big( \sum_{n \geq 1} L^{k n} \, t_{k n} \Big) \, C 
      \, \exp\Big( -\sum_{n \geq 1} R^{k n} \, t_{k n} \Big) = C \, ,
\ee
which shows that $\Phi^{(1)}$ and thus also $\Phi$ 
does not depend on the variables $t_{k n}$, $n=1,2, \ldots$. 
\hfill$\square$

\section{$\A$-modules, connections, and linear systems}
\label{section:A-modules}
\setcounter{equation}{0}
In this section we introduce a notion of left and right $\A$-module, 
where $\A$ is a WNA algebra. In terms of connections on such a module, 
we formulate a linear system whose integrability is a consequence of 
the nonassociative hierarchy (\ref{na_hier}).

\subsection{$\A$-modules}
\label{subsection:A-modules}
Let $\mathfrak{L}$ be an $\mathcal{R}$-module with a left action of a WNA algebra
$\A$ (over the commutative ring $\mathcal{R}$).
We call $\mathfrak{L}$ a \emph{left $\A$-module} if
\be
    a \, (b \,q) = (a b) \, q
    \qquad\quad
    \forall \,  a \in \A, \,  b \in \A', \, q \in \mathfrak{L} \; .
\ee
Note that, in general, $a \, (f \, q) \neq (a \, f) \, q$ for $f \not\in \A'$. 
Similarly, a \emph{right $\A$-module} $\mathfrak{R}$ means an $\mathcal{R}$-module
with a right action of $\A$ such that
\be
    (\bq \, b) \, a = \bq \, (b a)
    \qquad\quad
    \forall \, a \in \A, \, b \in \A', \, \bq \in \mathfrak{R} \; .
\ee
The algebra $\A$ itself is both a left and a right $\A$-module.
\vskip.2cm

Fixing some $f \in \A$, $f \not\in \A'$, we define recursively actions
$\circ_n$ via $a \circ_1 q = a \, q$, $\bq \circ_1 a = \bq \,a$, and
\be
    a \circ_{n+1} q = a \, (f \circ_n q) - (a f) \circ_n q \, , \qquad
    \bq \circ_{n+1} a = \bq \circ_n (f a) - (\bq \circ_n f) \, a \; .
\ee
They depend only on the equivalence class of $f$ in $\A/\A'$.
Several of our previous results (where the modules were given by $\A$ itself)
generalize to the present setting.
In particular, for $m,n=1,2,\ldots$ we obtain
\be
     a \circ_m (f \circ_n q) - (a \circ_m f) \circ_n q
 &=& a \circ_{m+n} q  \, ,  \\
     \bq \circ_m (f \circ_n a) - (\bq \circ_m f) \circ_n a
 &=& \bq \circ_{m+n} a \; .
\ee
\vskip.2cm

\noindent
\textbf{Example.} Let $\mathfrak{L}_{\mathrm{free}}$ be the \emph{left}
$\mathcal{A}_{\mathrm{free}}$-module (with $\mathcal{A}_{\mathrm{free}}$ defined 
in section~\ref{subsection:freeWNA}), freely generated by
elements $\mu_n,\,n=0,1,\ldots$, and $\mathfrak{R}_{\mathrm{free}}$ the \emph{right}
$\mathcal{A}_{\mathrm{free}}$-module, freely generated by elements 
$\bmu_n,\,n=0,1,\ldots$. They can be extended to left, respectively right, 
$\A_{\mathrm{free}}(f)$-modules by setting
\be
    f \, \mu_n = \mu_{n+1} \, , \qquad \quad  
    \bmu_n \, f = \bmu_{n+1}  \qquad \quad n=1,2,\ldots \; .
\ee
Hence
\be
    \mu_n = \hat{L}_f^n(\mu_0) \, , \qquad \bmu_n = \hat{R}_f^n(\bmu_0) \; . 
\ee
As a consequence, $\mathfrak{L}_{\mathrm{free}}$ and $\mathfrak{R}_{\mathrm{free}}$ 
are cyclic left, respectively right, $\A_{\mathrm{free}}(f)$-modules with 
generators $\mu_0$, respectively $\bmu_0$.
\hfill $\square$

\subsection{Connections} 
\label{subsection:connections} 
Let $\A$ be a WNA algebra and $\delta$ a derivation of $\A$.
A $\delta$-based \emph{module derivation} of a left $\A$-module $\mathfrak{L}$ 
is a linear map $\nabla \, : \, \mathfrak{L} \rightarrow \mathfrak{L}$,
with the property 
\be
    \nabla (a \, q) = \delta(a) \, q + a \, \nabla(q)
    \qquad \quad
    \forall \,  a \in \A \, , \,  q \in \mathfrak{L} \, .
\ee
Correspondingly, a $\delta$-based \emph{module derivation} of a right $\A$-module 
$\mathfrak{R}$ is a linear map $\nabla  \, : \, \mathfrak{R} \rightarrow \mathfrak{R}$, 
such that 
\be
    \nabla(\bq \, a) = \nabla(\bq) \, a + \bq \, \delta(a)
    \qquad \quad
    \forall \,  a \in \A \, , \,  \bq \in \mathfrak{R}  \; .
\ee
\vskip.2cm

The following propositions are proved by straightforward generalization 
of the arguments which led to the corresponding results in 
sections~\ref{section:derivations} and \ref{section:KPids}.

\begin{proposition}
\label{prop:nabla-circ_n} 
Let $\delta$ be a derivation of a WNA algebra $\A$ such that $\delta(\A) \subset \A'$. 
Any $\delta$-based module derivation of a left $\A$-module $\mathfrak{L}$, 
respectively a right $\A$-module $\mathfrak{R}$, is also a module derivation 
with respect to all actions $\circ_n$, $n=1,2, \ldots$, i.e. 
\be
   \nabla (a \circ_n q) = \delta(a) \circ_n q + a \circ_n \nabla(q)
   \qquad \quad
    \forall \,  a \in \A \, , \,  q \in \mathfrak{L} \, ,
\ee
respectively 
\be
   \nabla(\bq \circ_n a) = \nabla(\bq) \circ_n a + \bq \circ_n \delta(a)
    \qquad \quad
    \forall \,  a \in \A \, , \bq \in \mathfrak{R} \; . 
\ee
(The actions $\circ_n$ are defined in terms of an element $f \in \A$.) 
\hfill $\square$
\end{proposition}
\medskip

In the following, let $f \in \A \setminus \A'$ be such that the subalgebra $\A(f)$ 
generated by $f$ is $\delta$-compatible. Furthermore, let 
$\mathfrak{L}(f,q)$, respectively $\mathfrak{R}(f,\bq)$, be cyclic 
$\A(f)$-modules with generators $q$, respectively $\bq$. We further assume 
that these modules admit, for $n=1,2, \ldots$, a $\delta_n$-based module derivation
determined by
\be
    \nabla_n(q) := f \circ_n q \, , \qquad
    \nabla_n(\bq) := \bq \circ_n f  \, ,      \label{nabla_n-def}
\ee
respectively. We will refer to this property as \emph{$\nabla$-compatibility}. 

\begin{proposition}
\be
     [ \nabla_m , \nabla_n ] = 0  \qquad \quad m,n = 1,2, \ldots \; .
\ee
Calling the set $\{ \nabla_n \}_{n=1}^\infty$ of module derivations a 
\emph{connection}, this means that its curvature vanishes. 
\hfill $\square$
\end{proposition}
\bigskip

Let $\hH_n : \mathfrak{L}(f,q) \rightarrow \mathfrak{L}(f,q)$ and 
$\hE_n : \mathfrak{R}(f,\bq) \rightarrow \mathfrak{R}(f,\bq)$ be the linear maps 
recursively defined by $\hH_0(q) = q$, $\hE_0(\bq) = \bq$, and
\be
    \hH_{n+1}(q) = f \, \hH_n(q) \, , \qquad
    \hE_{n+1}(\bq) = \hE_n(\bq) \, f \, ,  \qquad n=0,1,2, \ldots  \; .   \label{mod-HE-recur}
\ee

\begin{proposition}
\be
    \hH_n = \chi_n(\tilde{\nabla}) \, , \qquad
    \hE_n = (-1)^n \chi_n(-\tilde{\nabla})  \qquad n=0,1,2, \ldots \, , 
\ee
with $\tilde{\nabla} := (\nabla_1, \nabla_2/2, \nabla_3/3, \ldots)$ and the elementary 
Schur polynomials $\chi_n$. 
\hfill $\square$
\end{proposition}
\bigskip

With the help of an indeterminate $\la$, the recursion relations (\ref{mod-HE-recur}) 
can be expressed as follows,
\be
    \hH(\la)(q) = q + \la \, f \, \hH(\la)(q) \, , \qquad
    \hE(\la)(\bq) = \bq + \la \, \hE(\la)(\bq) \, f \, ,    \label{mod-HlaEla-eqs}
\ee
with formal power series 
\be
     \hH(\la) := \sum_{n=0}^\infty \la^n \, \hH_n
              = \exp\Big( \sum_{n \geq 1} \frac{\la^n}{n} \, \nabla_n \Big)
        \, , \quad
     \hE(\la) := \sum_{n=0}^\infty \la^n \, \hE_n
              = \exp\Big( - \sum_{n \geq 1} \frac{(-\la)^n}{n} \, \nabla_n \Big)
                \; .
\ee
Since the $\nabla_n$ have the module derivation property and commute, the maps 
$\hH(\la)$ and $\hE(\la)$ satisfy 
\be
    \hH(\la)( a \circ_n q ) = H(\la)(a) \circ_n \hH(\la)(q) \, , \qquad
    \hE(\la)( \bq \circ_n a ) = \hE(\la)(\bq) \circ_n E(\la)(a) \, ,
\ee
with $H(\la), E(\la)$ defined in section~\ref{section:KPids}.
\vskip.2cm

$\hH(\la)$ has the inverse $\exp(-\sum_{n \geq 1} (\la^n/n) \, \nabla_n)$, 
which is the above expression for $\hE(-\la)$, but now built with the 
connection on $\mathfrak{L}$. Correspondingly, the defining formal power 
series for $\hH(-\la)$, now built with the connection on $\mathfrak{R}$, 
is the inverse of $\hE(\la)$. This simply means that we can consider 
$\hH(\la)$ and $\hE(\la)$ as being defined on both modules, $\mathfrak{L}(f,q)$ 
and $\mathfrak{R}(f,\bq)$, and they are related by $\hH(\la) \hE(-\la) = \mathrm{id}$. 
With this notation, we obtain 
\be
   \hH(\la)(\bq \, a) = \hH(\la)(\bq) \, H(\la)(a) \, , \qquad
   \hE(\la)(a \, q ) = E(\la)(a) \, \hE(\la)(q) \; .
\ee
Acting on the two equations (\ref{mod-HlaEla-eqs}) with $\hE(-\la)$, 
respectively $\hH(-\la)$, leads to 
\be
   \hE(\la)(q) = q + \la \, E(\la)(f) \, q \, , \qquad
   \hH(\la)(\bq) = \bq + \bq \, \la \, H(\la)(f) \, ,   \label{mod-hatEH(lambda)}
\ee
and thus
\be
    \hE_{n+1}(q) =  E_n(f) \, q \, , \qquad
    \hH_{n+1}(\bq) = \bq \, H_n(f)  \qquad
     n=0,1,\ldots  \; .       \label{mod-EHrecur}
\ee
 From these equations (for $n=1,2,\ldots$) the `KP identities' (\ref{E-KPmn}), 
respectively (\ref{H-KPmn}), are recovered as `integrability conditions' (see 
also \cite{DNS89}). 
\vskip.2cm

\noindent
\textbf{Example.} In the case of the free modules $\mathfrak{L}_{\mathrm{free}}$
and $\mathfrak{R}_{\mathrm{free}}$, for $n=1,2,\ldots$ we can consistently define 
a $\delta_n$-based module derivation by setting
\be
    \nabla_n (\mu_0) := f \circ_n \mu_0 \, , \qquad 
    \nabla_n(\bmu_0) := \bmu_0 \circ_n f  \, , 
\ee
respectively. The identities (\ref{mod-EHrecur}) then take the form
\be
    \hE_{n+1}(\mu_0) =  e_n \, \mu_0 \, , \qquad
    \hH_{n+1}(\bmu_0) = \bmu_0 \, h_n  \qquad
     n=0,1,\ldots  \; .
\ee
\hfill $\square$

\subsection{Linear systems and the potential KP hierarchy}
\label{subsection:mod-linsys}
Let $\A$ now be a WNA algebra over $\mathcal{R} = \mathcal{C}^\infty$ 
(the ring of smooth real or complex functions of $t_1,t_2,\ldots$). 
On $\mathfrak{L}(f,q)$, respectively $\mathfrak{R}(f,\bq)$, 
let us consider the linear systems
\be
    q_{t_n} = \nabla_n(q) \, , \qquad
    \bq_{t_n} = \nabla_n(\bq )  \qquad \quad n=1,2, \ldots  \, ,  \label{q-linsys} 
\ee
with the connections defined in (\ref{nabla_n-def}). 
By use of proposition~\ref{prop:nabla-circ_n}, each of these two systems is 
seen to be compatible if $f$ solves the nonlinear hierarchy (\ref{na_hier}). 
With the help of the identities (\ref{mod-hatEH(lambda)}), (\ref{q-linsys}) can 
be written as
\be
    q_{-[\la]} = q - \la \, f_{-[\la]} \, q \, , \qquad
    \bq_{[\la]} = \bq + \bq \, \la \, f_{[\la]} \, ,    \label{q-linsys2}
\ee
using the notation of section~\ref{subsection:KP}. 
Furthermore, formal solutions of these linear systems are obtained as follows.

\begin{proposition}
\label{prop:hatS}
Let $\{ \nabla_n \}_{n=1}^\infty$ be the connection defined on 
$\mathfrak{L}(f_0,q_0)$, respectively $\mathfrak{R}(f_0,\bq_0)$, 
according to (\ref{nabla_n-def}) (assuming $\delta$-compatible $\A(f_0)$ 
and $\nabla$-compatible modules). 
Let $f_0, q_0, \bq_0$ be constant (i.e., independent of $t_1,t_2,\ldots$). 
Then 
\be
    q := \hat{S}(q_0) \, , \quad
    \bq  := \hat{S}(\bq_0)
    \qquad \mbox{where} \quad
    \hat{S} := \exp\Big( \sum_{n \geq 1} t_n \, \nabla_n \Big) 
\ee
are formal solutions of (\ref{q-linsys}) with $f = S(f_0)$, where $S$ 
is defined in (\ref{S}). 
\end{proposition}
{\it Proof:} Using $\hat{S}(a \circ_n q_0) = S(a) \circ_n \hat{S}(q_0)$ for 
all $a \in \A(f_0)$, we have
\bez
  \pa_{t_n}(q) = \hat{S}(\nabla_n(q_0)) 
              = \hat{S}( f_0 \circ_n q_0 ) 
              = S(f_0) \circ_n \hat{S}(q_0)
              = f \circ_n q    \; .
\eez
Since $[f] = [f_0] \in \A(f_0)/\A(f_0)'$, the product $\circ_n$ is equivalently 
defined in terms of $f$ (instead of $f_0$), and we conclude that 
$\pa_{t_n}(q) = f \circ_n q$, which proves our assertion. Since
$\hat{S}(\nabla_n(q_0)) = \nabla_n(\hat{S}(q_0)) = \nabla_n(q)$, this also 
shows that $\nabla_n$ defined via (\ref{nabla_n-def}) in terms of $f_0, q_0$ 
coincides with $\nabla_n$ defined in terms of $f,q$. 
A corresponding argument applies to the right module linear system. 
\hfill $\square$
\bigskip

Recall from section~\ref{section:nahier} 
that $S(f_0)$ satisfies the nonassociative hierarchy (\ref{na_hier}). 
If there is a decomposition of $f$ of the form $f = \nu - \phi$ with constant 
$\nu$ and $\phi \in \A'$ (cf. (\ref{f_decomp})), the linear systems imply 
in particular $q_x = (\nu - \phi) \, q$ and $\bq_x = \bq \, (\nu - \phi)$, 
respectively, and then (\ref{q-linsys2}) can be written as
\be
   q_{-[\la]} = q - \la \, q_x - \la \, (\phi - \phi_{-[\la]}) \, q \, , \qquad
   \bq_{[\la]} = \bq + \bq_x \, \la - \bq \, \la \, (\phi_{[\la]} - \phi) \; . 
        \label{q-linsys3}
\ee
\vskip.2cm

\noindent
{\bf Remark.} Suppose $q,\bq$ are formal series in an indeterminate $\zeta$. 
Furthermore, let
\be
    q =  w \, e^{\xi(\zeta)} \, , \qquad 
    \bq = \bar{w} \, e^{-\xi(\zeta)} \, , \qquad
    \xi(\zeta) := \sum_{n \geq 1} t_n \, \zeta^n \, ,
\ee
where $w,\bar{w}$ are formal series in $\zeta^{-1}$ of the form 
$w = \sum_{n \geq 0} w_n \, \zeta^{-n}$, 
$\bar{w} = \sum_{n \geq 0} \bar{w}_n \, \zeta^{-n}$ with 
constant $w_0, \bar{w}_0 \neq 0$. Using
\be
    \exp\Big(-\sum_{n \geq 1} \frac{1}{n} (\la \zeta)^n \Big) = 1 - \la \, \zeta 
\ee
(as a formal power series), we obtain
\be
     (1-\la \, \zeta)(w_{-[\la]} - w) 
 &=& - \la \, w_x + \la \, (\phi_{-[\la]} - \phi) \, w \, , \label{mod-ex-BA} \\
     (1 - \la \, \zeta)(\bar{w}_{[\la]} - \bar{w}) 
 &=& - \la \, \bar{w}_x + \la \, \bar{w} \, (\phi_{[\la]} - \phi) \; .  \label{mod-ex-BA2}
\ee
 For $\la = \zeta^{-1}$, these equations reduce to
\be
    w_x = (\phi_{-[\zeta^{-1}]} - \phi) \, w  \, , \qquad
    \bar{w}_x = \bar{w} (\phi_{[\zeta^{-1}]} - \phi) \; .
\ee
Representing $\A'$ by a commutative algebra of functions, and correspondingly 
for $\mathfrak{L}(f,q)$ and $\mathfrak{R}(f,\bq)$, with 
$\phi \mapsto \tau_x/\tau$ the last equations integrate to
\be
    w = {\tau_{-[\zeta^{-1}]} \over \tau} \, w_0 \, , \qquad
    \bar{w} = \bar{w}_0 \, {\tau_{[\zeta^{-1}]} \over \tau} \, ,
\ee
which are familiar formulae for the Baker-Akhiezer function of the scalar KP hierarchy, 
and its adjoint (cf. \cite{Dick03}, for example). 
By use of these expressions, (\ref{mod-ex-BA}), respectively (\ref{mod-ex-BA2}), 
becomes equivalent to 
\be
   ( \la^{-1} - \zeta ) \, \left( \tau_{[\la]+[\zeta^{-1}]} \, \tau
         - \tau_{[\la]} \, \tau_{[\zeta^{-1}]} \right)  
   = \tau_{[\la]} \, \tau_{[\zeta^{-1}],x} 
         - \tau_{[\la],x} \, \tau_{[\zeta^{-1}]} \, ,
\ee
which is the differential Fay identity (\ref{diffFay}).
\hfill $\square$
\vskip.2cm

\noindent
{\bf Example.} The following supplements the material in section~\ref{subsection:KPsol}. 
Let $\mathcal{M}(M,N)$ be the algebra of $M \times N$ matrices, 
taken over $\mathcal{C}^\infty$, with the product (\ref{K-product}) which 
involves an  $N \times M$ matrix $K$.
Then $\K^M$ (where $\K$ stands for $\R$ or $\C$) extends to a left 
$\mathcal{M}(M,N)$-module by setting
\be
    A \ci \mu := A \, K \, \mu  \qquad \quad
          \forall A \in \mathcal{M}(M,N) , \, \mu \in \K^M \; .
\ee
Correspondingly, $\K^N$ extends to a right $\mathcal{M}(M,N)$-module
by setting
\be
    \bmu \ci A := \bmu^T \, K \, A  \qquad \quad
          \forall A \in \mathcal{M}(M,N), \, \bmu \in \K^N \; .
\ee
Let us introduce new actions by 
\be
    A \circ_1 \mu &:=& (A R) \ci \mu - A \ci (L\mu) = A \, (R K-K L) \, \mu \, , \\
    \bmu \circ_1 A &:=& (\bmu^T R)^T \ci A - \bmu \ci(L A) = \bmu^T (R K - K L) \, A \, , 
\ee
where $L,R$ are the matrices in section~\ref{subsection:KPsol}. 
Together with $\nu \circ_1 \mu := L \mu$ and $\bmu \circ_1 \nu := - \bmu^T R$, this 
turns the above modules into a left, respectively right $(\A,\circ_1)$-module. 
It is easily verified that
\be
    \begin{array}{lcl}
    \nu \circ_n \mu = L^n \mu \, , & \quad & \bmu \circ_n \nu = -\bmu^T R^n \, , \\
    A \circ_n \mu = A \, (R^n K - K L^n) \, \mu \, , & & 
           \bmu \circ_n A = \bmu^T (R^n K - K L^n) \, A \; .
    \end{array}
\ee
In the case under consideration, the linear systems (\ref{q-linsys2}) take the form 
\be
    q_{t_n} &=& f \circ_n q = (I_M + \phi \, K) L^n \, q - \phi \, R^n K \, q \, , \\
    \bq_{t_n}^T &=& \bq \circ_n f = -\bq^T R^n \, (I_N + K \, \phi) + \bq^T K \, L^n \phi \, , 
\ee
which are solved by
\be
      q &=& (I_M + \phi \, K) \, e^{\xi(L)} \, \mu_0 
         = e^{\xi(L)} \, (I_M - B)^{-1} \, \mu_0 \, , \\
  \bq^T &=& \bmu_0^T e^{-\xi(R)} \, (I_N + K \, \phi) 
         = \bmu_0^T (I_N - \bar{B})^{-1} \, e^{-\xi(R)} \, , 
\ee
where $\mu_0 \in \mathbb{K}^M$, $\bmu_0 \in \mathbb{K}^N$, 
$\phi$ is the solution of (\ref{nahier-phi-ci}) obtained in section~\ref{subsection:KPsol}, 
and $B = C e^{-\xi(R)} K e^{\xi(L)}$, $\bar{B} = e^{-\xi(R)} K e^{\xi(L)} C$. 
Then $q, \, \bq$ also satisfy 
\be
   q_{-[\la]} = q - \la \, q_x - \la \, (\phi - \phi_{-[\la]})(R K - K L) q
       \, , \; \; 
   \bq^T_{[\la]} = \bq^T + \la \, \bq^T_x 
      - \la \, \bq^T (R K -L K)(\phi_{[\la]} - \phi) \, . \;
\ee
Choosing $u \in \mathbb{K}^M$ and $v \in \mathbb{K}^N$ such that $R K - K L = v u^T$
(cf. section~\ref{subsection:KPsol}), it follows that $\psi := u^T q$ and 
$\bar{\psi} = \bq^T v$ solve the scalar equations
\be
   \psi_{-[\la]} = \psi - \la \, \psi_x
      - \la \, (\varphi - \varphi_{-[\la]}) \, \psi \, , \qquad
   \bpsi_{[\la]} = \bpsi + \la \, \bpsi_x 
      - \la \, \bpsi \, (\varphi_{[\la]} - \varphi) \, ,
\ee
where $\varphi := \Psi(\phi) = u^T \phi \, v$, with the solution $\phi$ from 
section~\ref{subsection:KPsol}. The last equations do not make reference 
to $L$ and $R$ (and thus the WNA structure) any more, they constitute linear 
systems of the scalar potential KP hierarchy with dependent variable $\varphi$.
\hfill $\square$

\section{Conclusions}
\label{section:concl}
\setcounter{equation}{0}
This work originated from our recent study \cite{DMH05KPalgebra} of the combinatorics 
underlying the building rules of KP hierarchy equations, which led to a quasi-shuffle 
algebra. In \cite{DMH05KPalgebra} we used a
weakly nonassociative extension of an associative algebra as a technical
sidestep in order to simplify certain calculations. 
The present work shows that there are in fact deep relations between the 
KP hierarchy and (weakly) nonassociative algebras. 
\vskip.2cm

In particular, we have shown that any solution of the `nonassociative hierarchy' 
(\ref{naKP_hier}) of ordinary differential equations in any WNA algebra $\A$ 
leads to a solution of the KP hierarchy, with dependent variable in the 
associative algebra $\A'$. In special cases (i.e., for certain classes of WNA algebras) 
we can explicitly construct such solutions, as done in section~\ref{section:solutions}. 
But in general the nonassociative hierarchy (\ref{naKP_hier}) apparently 
cannot be solved explicitly.

Though we do not know yet to what extent solutions of the `nonassociative hierarchy' 
(\ref{naKP_hier}) really exhaust the set of solutions of KP hierarchies, 
we have shown that a significant subset of solutions is reached in this way, 
including the multi-soliton solutions of the scalar KP hierarchy. 
Keeping a possible restriction to a subset of solutions in mind, 
the nonassociative hierarchy (\ref{naKP_hier}) achieves to 
decouple the partial derivatives $\phi_{t_n}$ in the potential KP hierarchy, 
and presents its content in terms of a hierarchy of \emph{ordinary} 
differential equations. We should mention that systems of ordinary Riccati 
type equations related in such a way to the scalar KP hierarchy already appeared 
in \cite{DNS89,Haak96,FMP98}. 
\vskip.2cm

It would be of great interest to explore the relations with other 
methods to construct solutions of the KP hierarchy, in particular with the 
famous ones of the Japanese group \cite{Sato81,Sato+Sato82,DKJM83} 
and of Segal and Wilson \cite{Sega+Wils85}.
\vskip.2cm

Recalling the construction of commuting derivations of a nonassociative algebra, 
generated by a single element $f$, as outlined in the introduction, 
the question remains whether there are hierarchies associated with 
other (not weakly) nonassociative algebras. 
Perhaps there is a nonassociative algebra related to other integrable 
hierarchies in a similar way. At least, we expect that the discrete KP 
\cite{Adle+vanM99,Kupe00} and two-dimensional Toda lattice hierarchy 
\cite{Ueno+Taka84} get a place on this stage. 
\vskip.2cm

In view of the role which integrable systems, and in particular 
the KP hierarchy, play in algebraic and differential geometry, one should expect 
corresponding nonassociative versions to appear in nonassociative generalizations 
of (noncommutative) geometries (see \cite{Mack+Scho92b,Sabi00,Nest+Sabi00a,Wein00,Kiny+Wein01,Ione03,Maji05nagt}, 
for example). 
We also refer to \cite{LPS98} for an overview of some of the uses of nonassociative 
algebras in mathematics and physics.

\end{document}